\documentclass[prd,reprint,showpacs,showkeys,shortbibliography]{revtex4-1}

\usepackage[colorlinks = true,
            linkcolor = blue,
            urlcolor  = blue,
            citecolor = blue,
            anchorcolor = blue]{hyperref}

\usepackage{graphicx}
\usepackage{amsfonts}
\usepackage{amsmath}
\usepackage{amssymb}
\usepackage[squaren]{SIunits}

\newcommand{\um}[1]{\ensuremath{\,\mathrm{#1}}}    % units of measure
\newcommand{\di}{\ensuremath{\,\mathrm{d}}}        % differential
\newcommand{\vect}[1]{\ensuremath{\mathbf{#1}}}     % 3D vector

\newcommand{\Mn}{\ensuremath{m_{\text{n}}}} % neutron mass, 939.565413(6) MeV
\newcommand{\Mp}{\ensuremath{m_{\text{p}}}} % proton mass, 938.272081(6) MeV
\newcommand{\Ek}{\ensuremath{E_{\text{k}}}} % kinetic energy

\begin{document}

\title{Neutron astronomy}

\author{Diego Casadei}
\email{diego.casadei@cern.ch}

\affiliation{
	School of Physics and Astronomy, University of Birmingham\\
	and School of Engineering, FHNW
}

% \altaffiliation[Correspondence to:]{
%   FHNW, School of Engineering, i4Ds, 5.2C01
%   Bahnhofstrasse 6, 5210 Windisch, Switzerland
% %  Phone: +41-56-202-7684
% }

\date{9 Jan 2017.  Last revision: \today}

\begin{abstract}
  Neutrons travel along straight lines in free space, but only survive
  for a distance which depends on their energy. Thus, detecting
  neutrons in space in principle provides directional and distance
  information. Apart from secondary neutrons produced by cosmic-ray
  interactions in the Earth atmosphere, which are the dominant
  background, direct neutron emission is caused by solar flares, with
  clear time correlation with X-rays, which can be measured from few
  tens MeV up to few GeV. There is no detectable astrophysical source
  up to the PeV scale, when neutrons coming from supernova remnants
  may reach the Earth before decaying.  In addition, ultra high energy
  neutrons are the most plausible explanation for the measured
  anisotropy of cosmic-ray showers produced in the atmosphere above
  $10^{18}$~eV. From the GeV to the PeV scale, the expected neutron
  flux is very low and not too different from the antiproton flux, as
  the same cosmic-ray collisions with the interstellar medium which
  can produce antiprotons can also produce neutrons and antineutrons.
  This background flux of cosmic-ray neutrons is very low and has not
  yet been detected.  Measuring the neutron energy spectrum in space
  is a very effective way of searching for decays of exotic
  particles. For example, dark matter could consist of Weakly
  Interacting Massive Particles (WIMPs), which may annihilate into
  final states with particle and antiparticle pairs.  Consequently, a
  number of indirect WIMP searches are being carried on, focusing on
  positron and antiproton spectra.  However, no experiment is
  presently foreseen to look for ``bumps'' in the neutron energy
  spectrum, which is virtually background free. Here we consider the
  implications of a measurement of the neutron energy spectrum in
  astronomy and astrophysics and list the interesting energy regions
  in the search for WIMPs.
\end{abstract}

\keywords{cosmic rays; neutrons; dark matter}

\pacs{95.30.Cq, % Elementary particle processes
      95.35.+d, % Dark matter
      95.55.-n, % Astronomical and space-research instrumentation
      95.55.Vj  % Neutrino, muon, pion, and other elementary particle detectors; cosmic ray detectors
}

%%%%%%%%%%%%%%%%%%%
%%%%%%%%%%%%%%%%%%%
\maketitle
%%%%%%%%%%%%%%%%%%%
%%%%%%%%%%%%%%%%%%%

%%%%%%%%%%%%%%%%%%%%%%%%%%%%%%%%%%%%%%%%%%%%%%%%%%%%%%%%%
\section{Introduction}
%%%%%%%%%%%%%%%%%%%%%%%%%%%%%%%%%%%%%%%%%%%%%%%%%%%%%%%%%

 Cosmic rays (CR) have been studied since one century with experiments
 performed on the ground or underground, on high mountains, on
 stratospheric baloons, and on satellites.  They consist mainly
 of protons ($\sim90\%$), helium nuclei and electrons, although all
 elements are also present (see for example \cite{longair}).
 Although the magnetic field intensities are usually very weak in empty
 space (with the important exceptions of the sites where charged CR
 get accelerated), the distances traveled between their sources and
 the Earth are so big that the deviations induced by the magnetic
 fields completely erase the information about the location of CR
 sources: no astronomy is possible with charged particles.

 On the other hand, neutral particles travel along straight lines,
 hence provide directional information on the location of their
 sources.  This is the case for photons and neutrinos, for example.
 Photons can be detected over an enormous energy range, and are the
 classical carriers for astronomy.  They even allow us to achieve
 3-dimensional maps, by exploiting the cosmological redshift to date
 their sources back to the recombination period.
 In the past few decades a big progress has been achieved on neutrino
 astrophysics, and huge detectors (necessary because of the very small
 neutrino cross-section) are now operating in different locations all
 around the world.
%
 % Very low-energy neutrinos cannot be directly detected.  If that were
 % possible, one could get a picture of the visible universe at an
 % earlier stage than the recombination epoch, as neutrinos decoupled
 % from the thermal equilibrium earlier than photons.

 Among the known particles, neutrons possess very interesting
 characteristics.  They are neutral like photons and neutrinos, hence
 they travel along straight lines.  However, they have a long but
 finite life time ($\tau_0 = 881.5\pm1.5$~s \cite{pdb2016}) hence
 travel for a typical distance of $\gamma c\tau_0$ (more details in
 section~\ref{sec-propagation} below), where
 $\gamma = (1 - v^2/c^2)^{-1/2}$ is the Lorentz factor and
 $c\tau_0 = 2.64 \times 10^{11} \um{m} = 1.77 \um{AU} = 8.6 \times
 10^{-6} \um{pc}$.
 Hence there is a typical energy threshold for the neutrons coming
 from a source at some distance $x$.  Furthermore, most astrophysical
 sources produce energy spectra which are steeply falling down
 functions, hence the expected neutron energy distribution may be
 detectable in a relatively narrow range.  This means that neutrons at
 different energies probe different depths, creating a sort of
 ``astro-tomography'' of the environment around the observer.

 Neutrons from the Sun need to be above about 20 MeV to survive until
 the Earth and their spectrum steeply falls down at higher energies,
 such that they are detectable up to several hundred MeV.  For larger
 energies, there is no standard source of neutrons up to the PeV
 region, where neutrons emitted during supernova explosions may reach
 the solar system.
 Thus there is a very wide range, spanning many decades in energy, in
 which the expected flux of neutrons is very low and so far remained
 undetected.  This is essentially a background-free region for the
 search for new particles, for example the Weakly Interacting Massive
 Particles (WIMPs), which are among the favored dark matter
 candidates.

 At present, several space experiments (like AMS-02 and DAMPE, for
 example) are looking for the signatures of WIMP decays in the 
 proton-antiproton and electron-positron final states.  As a
 neutron-antineutron final state is essentially equivalent to a
 proton-antiproton decay (both come from hadronization of a
 quark-antiquark pair), it comes out that WIMP searches may also be
 carried on by focusing on the energy spectrum of cosmic neutrons.
 The advantage is that the only source of instrumental background is
 connected to the cosmic gamma rays and to the secondary neutrons
 produced in the spacecraft and in the Earth atmosphere: an
 anticoincidence shield is sufficient to get rid of all charged
 particles (in particular of CR protons).  By exploiting pulse shape
 discrimination techniques, the gamma-ray background can be easily
 suppressed.  In addition, a directional neutron detector can easily
 get rid of the atmospheric background (alternatively one could
 perform a differential measurement between upward and downward going
 neutrons), leaving only the diffuse cosmic neutrons as background
 sources.
 The flux of cosmic neutrons is so low that one is left almost with a
 background-free search for faint signals.

 In the rest of the paper, the known sources of cosmic neutrons are
 reviewed (section~\ref{sec-sources}) after some details of neutron
 propagation are clarified (section~\ref{sec-propagation}).  Next the
 interesting energy regions for dark matter particles are reviewed
 (section~\ref{sec-DM}).  Finally, neutron detection techniques are
 studied, which can be exploited in the search for WIMP decays
 (section~\ref{sec-detection}).

%%%%%%%%%%%%%%%%%%%%%%%%%%%%%%%%%%%%%%%%%%%%%%%%%%%%%%%%%
\section{Propagation}\label{sec-propagation}
%%%%%%%%%%%%%%%%%%%%%%%%%%%%%%%%%%%%%%%%%%%%%%%%%%%%%%%%%

 For a neutron with energy $E = \gamma \Mn c^2$, with the neutron mass
 $\Mn = 939.565 379 (21)$ MeV$/c^2$ \cite{pdb2016}, the time needed to
 travel the distance $x$ is
 \begin{equation}
   \label{eq-delay}
   \Delta t(x) = \frac{x}{v} 
               = \frac{x}{c} \frac{\gamma}{\sqrt{\gamma^2 - 1}}
 \end{equation}
 when measured in the observer's reference system, while the
 corresponding neutron proper time is
 \begin{equation}
   \label{eq-proper-delay}
   \tau(x) = \frac{\Delta t(x)}{\gamma} 
           = \frac{x}{c} \frac{1}{\sqrt{\gamma^2 - 1}} 
 \end{equation}
 assuming that the free neutron is created at the moment of its
 emission from the source.
 The life time of a free neutron is $\tau_0$, hence the survival
 probability is $P_\text{surv}(x) = \exp(-\tau/\tau_0)$, or
 \begin{equation}
   \label{eq-P-no-decay}
   P_\text{surv}(x) = \exp \left[ - \dfrac{x}{c\tau_0 \sqrt{\gamma^2 - 1}} \right]
 \end{equation}

 Neglecting interactions (see below), \eqref{eq-P-no-decay} also gives
 the fraction of monoenergetic neutrons surviving after the distance
 $x$.  Unless $\gamma$ is small, to a good approximation for a source
 at distance $2x$ one gets the same fraction of surviving neutrons at
 double energy.  If one sets an energy threshold such that, say,
 $P_\text{surv}(x) \ge 0.01$, then this threshold is proportional to
 $x$ for $\gamma\gg1$.  Similarly, the typical traveled distance for
 $\gamma\gg1$ is $\gamma c\tau_0$, which scales linearly with the
 neutron energy.
 Thus neutrons at different energies probe different depths, because
 they cannot come from distances much smaller than $c\tau_0$.  In this
 sense, the knowledge of their direction and energy can be in
 principle exploited to perform a sort of ``astro-tomography'' of the
 environment around the observer.

 Neutrons propagating in the interstellar medium or through stellar
 atmospheres may undergo hadronic interactions.  The cross-section for
 n-p interactions on target hydrogen (the most abundant species) is
 about 1.4 barns at 100 MeV and decreases to a minimum just above
 30~mb at 1 GeV, then gently increases up to about 40~mb at 2--3 GeV,
 remaining at this level up to several hundred GeV (see Figure 49.9 of
 \cite{pdb2016}).  Above the minimum, inelastic interactions are the
 majority, whereas below about 1~GeV one has elastic collisions.

 The mean interaction length for neutrons is
 $\lambda \approx 60 \um{g} \um{cm}^{-2}$ and the probability of no
 collision $P_\text{nc} = \exp [-C \rho x/\lambda]$ has an
 energy-dependent coefficient $C$ which is proportional to the neutron
 cross-section.  Above 1 GeV, $C$ is a very weak function of the
 energy, such that $P_\text{nc}$ only depends on the distance $x$
 traveled through the medium with average density $\rho$.  At lower
 energies one has a rapidly increasing probability of performing
 elastic collisions.
 The latter decrease the neutron energy and induce a spread over the
 initial source direction, with RMS angular dispersion of about
 $\sqrt{n} \, 6^\circ$, where $n$ is the number of elastic collisions
 \cite{pdb2016}.  Hence neutron astronomy is only feasible when the
 expected number of collisions is smaller than one.  This is certainly
 the case for the propagation through the very low density
 interstellar medium, as $\lambda$ is ten times higher than the
 grammage traversed by charged CR during their \emph{diffusion} in the
 Galaxy, estimated from the measured ratio of secondary-to-primary
 nuclei.  The actual grammage encountered by neutrons is much smaller
 than the amount of matter traversed by charged particles, because the
 former travel along straight lines while the latter are significantly
 deflected by magnetic fields and perform a random motion inside the
 Galaxy.
 Thus, apart from very low energy neutrons (which typically decay
 before reaching the observer) the expected number of collisions is
 much smaller than one.  In this case the probability of collision
 $1-P_\text{nc}$ is very small and the fraction of lost neutrons can
 be well approximated by $C \rho x/\lambda$ (which is very small too,
 and can be neglected to first approximation).

 We can safely assume that inelastic collisions will produce a loss of
 neutrons, which adds up to the loss due to their beta decay.  Because
 elastic scattering deflect neutrons, a detector pointing toward a
 point source will most likely miss elastically scattered neutrons
 even in case of a single collision.  Thus we can assume that also
 elastic scattering produces a loss of neutrons (those originally
 traveling along a different direction, which reach the Earth after
 being deviated in an elastic collision, are ignored).  Hence the
 survival probability is obtained by correcting \eqref{eq-P-no-decay}
 for the losses due to collisions:
 \begin{equation}
   \label{eq-P-survival}
   \begin{split}
     P_\text{surv}(x) &= \exp \left[ - \dfrac{x}{c\tau_0 \sqrt{\gamma^2 - 1}}
                       - \frac{C x \rho}{\lambda} \right]
   \\
       &\simeq  \exp \left[ - \dfrac{x}{c\tau_0 \gamma}
                       - \frac{C x \rho}{\lambda} \right]
   % \\
   %     &\simeq \frac{C x \rho}{\lambda}
   %             \exp \left[ - \dfrac{x}{c\tau_0 \gamma} \right]
   %%%  this assumes also that $\lambda \gg C x \rho$  %%%
   \end{split}
 \end{equation}
 where the second expression is valid for $\gamma \gg 1$.  This
 condition is necessary if $x \gg c\tau_0$, in order to have a non
 negligible $P_\text{surv}(x)$ when $x$ is large.  In practice the
 approximation holds whenever neutrons are detected from a source
 located outside the solar system.

 Incidentally, note that the effects of elastic collisions depend on
 the goal of the analysis.  In \eqref{eq-P-survival} we consider the
 propagation through the interstellar medium.  However if one is
 interested into the propagation in the Earth atmosphere then elastic
 scattering changes the neutron energy but does not decrease their
 number \cite{Shibata_1994}.

 Consider the detection at time $t$ of neutrons with energy
 $E = \gamma \Mn c^2$ from a source at distance $x$, and let
 $I(\gamma, t; x)$ denote the measured flux in the solid angle
 $\di\Omega$ along the line of sight of the source.  The number of
 measured neutrons in $[\gamma, \gamma + \di\gamma]$ and
 $[t, t+\di t]$ is then $N(\gamma, t; x) = I(\gamma, t; x) \di\Omega$.
 The corresponding emitted flux at the source is
 $I(\gamma, t-\Delta t; 0)$, where the delay $\Delta t$ is given by
 \eqref{eq-delay} as a function of $x$ and $\gamma$, and the neutrons
 emitted by the source in the solid angle $\di\Omega$ along the line
 of sight are given by
 $N(\gamma, t-\Delta t; 0) = I(\gamma, t-\Delta t; 0) \di\Omega$.

 Our goal is to estimate $N(\gamma, t-\Delta t; 0)$ from the
 measurement of $N(\gamma, t; x)$.
 Once this is achieved, one can get the source term
 $Q(\gamma, t'; 0) = \frac{4\pi}{\di\Omega} N(\gamma, t'; 0)$, which
 describes the injected spectrum of neutrons as a function of time
 (denoted $t'$ to keep it different from the measurement time $t$).
 The measured number of neutrons is actually given by $N(\gamma, t;
 x)$ plus the background counts:
 \begin{equation}
   \label{eq-observed}
   N(\gamma, t; x) = e^{ - \frac{x}{c\tau_0 \sqrt{\gamma^2 - 1}}
                                 - \frac{C x \rho}{\lambda} }
                     N(\gamma, t-\Delta t; 0)
                   + \text{bkg}
 \end{equation}
 The latter may be due to instrumental effects, like internal
 radioactivity of the materials used to build the detector, and to
 sources of neutrons other than the observed one, like cosmic-ray and
 atmospheric neutrons.  Hence it is important to know this
 ``background'' flux as a function of energy.

%%%%%%%%%%%%%%%%%%%%%%%%%%%%%%%%%%%%%%%%%%%%%%%%%%%%%%%%%
\section{Known sources of neutrons}\label{sec-sources}
%%%%%%%%%%%%%%%%%%%%%%%%%%%%%%%%%%%%%%%%%%%%%%%%%%%%%%%%%

 We will take the following sources of neutrons into consideration.
 Three are localized: solar flares, supernova remnants, and a possible
 giant black hole at the Galaxy center.  They can be considered point
 sources, as even the closest ones, solar flares, have size of 1
 arcmin or less, much smaller than the angular resolution of any
 conceivable neutron detector at the relevant energies.  Two are
 diffuse sources: neutrons from CR interactions in the Galaxy and in
 the Earth atmosphere.  Localized sources are discussed first.

%%%%%%%%%%%%
\subsection{Solar flares}\label{sec-flares}
%%%%%%%%%%%%

 Solar flares are the most powerful events in the solar system,
 releasing $10^{25}$--$10^{26}$~J in few minutes \cite{Benz_2008}.  A
 large fraction of this energy, initially stored in magnetic fields in
 the solar corona, goes into the acceleration of ions and electrons.
 The latter emit X-rays by bremsstrahlung, hence are directly
 observable.  Indeed, solar flares are the brightest sources of X-rays,
 with energy ranging from few keV (dominated by thermal emission in the
 coronal region of the reconnected magnetic flux loop) to several
 hundreds keV.  Above 10--20 keV their distribution is clearly
 non-thermal, revealing the details of the parent distribution of the
 accelerated electrons, and is dominated by the footpoints of the
 magnetic loop, where downward electrons impact on the chromosphere,
 whose density is about 2 orders of magnitude larger than in the solar
 corona.

 As a consequence of a solar flare, neutrons can be detected on the
 Earth either as direct emission from the Sun or as secondary products
 in the atmosphere.  Their time signature is clearly different, as the
 secondary neutrons are produced by the accelerated protons, which
 reach the Earth only when the latter is magnetically connected to the
 flaring region.  Thus protons need to follow curved magnetic field
 lines and reach the Earth tens of minutes after the prompt phase of
 X-ray emission is detected.  Entering the Earth atmosphere, they
 produce showers of secondary particles, including additional
 neutrons.  Solar flares can be detected by the neutron monitors
 typically located on high mountains, by stratospheric balloon
 experiments, and by satellite instruments in low Earth orbit.

 If neutrons are emitted in the impulsive phase of a solar flare,
 those with highest energy arrive at the Earth $O(10^3)$~s after their
 emission (with at most a small delay with respect to X-rays), followed
 by less energetic neutrons.  Their flux should decrease with a
 typical decay time of $O(10^3)$~s \cite{Forrest_1969}.  Hence they
 are typically well separated in time from the secondary neutrons
 produced in the hadronic showers induced by the accelerated protons.

 During quiet Sun periods, there is no significant direct flux of
 solar neutrons.  Early attempts to measure such a flux date back to
 the sixties.
 \citet{Hess_1967} found no evidence for a diurnal neutron rate
 variation in 1962, using BF$_3$ counters on OSO-1 to look for solar
 neutrons between 10 keV and 10 MeV in a period without solar flares.
% A 1\% charged-particle rate variation would have been detectable.
 They have set an upper bound of $2\times10^{-3}$ Hz/cm$^2$ on solar
 neutrons within this energy range, more stringent than the result
 obtained at the same solar minimum by the Vela satellites
 \cite{Bame_1966}.

 \citet{Hess_1967} also emphasized that thermonuclear reactions in the
 solar atmosphere seem to yield a negligibly small neutron flux also
 in connection with solar flares, concluding that the biggest source
 of neutrons is from solar protons interacting in the Earth
 atmosphere.
 % As one expects an intensity peak after one interaction mean free path
 % ($\lambda \approx 60 \um{g}/\um{cm}^2$), balloon-borne (and low-Earth
 % satellite) neutron detectors should see a peak in the count rate
 % somewhat before sunset, when neutrons grazing the atmosphere pass
 % through $\approx 60 \um{g}/\um{cm}^2$ before reaching the detector.

 Direct observations of solar neutrons could only be made two decades
 later.  \citet{Chupp_1982} reported 50--600 MeV neutrons detected at
 the Earth in correspondence with the solar flare of 1980-06-21 01:18
 UT, for about 17 minutes.  This measurement was performed by the
 Gamma-Ray Spectrometer (GRS) flown on the \emph{Solar Maximum
   Mission} (SMM) and provided evidence for the acceleration of
 protons up to GeV energies during a solar flare.  The peak counting
 rate was $(3.8\pm0.6)\times10^{-2}$ neutrons/(cm$^2$ s) at about 130
 MeV.
 Neutrons have been detected also during the 1980-06-03 11:33 UT
 flare \cite{McDonald_1985,Chupp_1987} and during the 1982-06-03
 11:43 UT flare \cite{Debrunner_1983,Efimov83,kudela90}.   This was the first
 observation of solar neutrons from ground --- actually high
 mountains (Jungfraujoch and Lomnicky Stit) --- detectors.

 The time-dependent neutron flux from a short-duration ($\lesssim100$
 s) solar flare is a sensitive measure of the energy spectrum and
 number of flare-accelerated protons and nuclei, as shown by
 \citet{Ramaty_1983}.
 By looking at the protons from the decay of solar flare neutrons,
 \citet{Evenson_1983} could determine the spectrum of neutrons emitted
 during the 1980-06-21 solar flare in the range 10-100 MeV, by solving
 the diffusion equation under the assumption that neutrons and
 gamma-rays were generated simultaneously.

 \citet{Murphy_1987} emphasized that the same nuclear reactions which
 produce pions (detected in the gamma-ray range) also produce
 secondary protons and neutrons.  Looking at the gamma-ray and neutron
 observations from the 1982-06-03 flare, they have built a model based
 on thick-target interactions, in which nuclear reactions occur as the
 accelerated particles slow down and stop in the chromosphere.  They
 considered in details the production of neutrons from p-p, p-$\alpha$
 and $\alpha$-$\alpha$ interactions, neglecting heavier nuclei,
 obtaining a spectrum with a pronounced bump between 500 and 600 MeV.
 They identified two acceleration phases, with most gamma-rays at
 2.223 MeV and 4.1--6.4 MeV emitted during the first one, while the
 second phase is responsible for most neutrons detected at the Earth.

 Neutron monitors on the ground and in the mountains can also detect
 neutrons emitted from solar flares, although most neutrons are
 secondary products of solar or galactic CR interactions in the
 atmosphere.  The flux of charged particles depends on the location of
 the detector, because the Earth magnetic field acts as a shield
 against soft ions.
 What matter is the magnetic rigidity $R=pc/(Ze)$, where $p$ is the
 relativistic momentum and $Ze$ is the ion charge, as it is
 proportional to the curvature radius $r = R \sin\theta / (Bc)$, where
 $B$ is the magnetic field intensity and $\theta$ is the angle between
 the ion momentum and the magnetic field.  Most published results
 provide the CR flux as a function of the kinetic energy
 \begin{equation}
   \label{eq-kin-ene}
   \Ek = m c^2 [ \sqrt{1 + (ZeR)^2 / (m c^2)^2} - 1 ]
 \end{equation}
 When $R$ is measured in GV, \Ek\ is measured in GeV.

 \citet{Shea_1991} reported an increase of neutron rate detected by 7
 neutron monitors in USA as a consequence of the solar flare on
 1990-05-24 20:46 UT.  The detection at stations with geomagnetic
 cutoff rigidity of about 8 GV was interpreted as a proof that protons
 have been accelerated at least up to 7 GeV, before they interacted
 with ions in the solar atmosphere producing neutrons by spallation (a
 more detailed analysis of the same event is presented by
 \citet{Debrunner_1997}).
%
 % A more detailed analysis of the same event is presented by
 % \citet{Debrunner_1997}, who concluded that a first impulsive phase
 % with production of gamma-rays and neutrons above 200 MeV was followed
 % by a slower phase, with production of high-energy protons for about
 % 20 minutes, whose interactions fave pions and high-energy neutrons.
 % Finally, a third phase coincided with the release of protons in the
 % interplanetary space, starting after the onset of the event but
 % before the end of the second phase.
%
 In the same solar cycle, \citet{Muraki_1992} reported the detection
 of 50--360 MeV neutrons associated with the large flare on 1991-06-04
 03:37 UT, with a neutron telescope and a muon telescope located at
 Mount Norikura Cosmic Ray Laboratory at 2770 m above sea level.

 The problem with neutron monitors is that they cannot determine
 whether the neutron is primary or secondary, and they only detect the
 neutrons after they have undergone collisions in the atmosphere.
 Hence there is little information about the primary neutron energy.
 \citet{Shibata_1994} simulated the propagation of solar neutrons
 through the Earth atmosphere with Monte Carlo methods, finding that
 elastic scattering plays an important role at neutron energies below
 200 MeV.

 Another way of detecting solar neutrons is to focus on the
 enhancement of protons and electrons resulting from their beta decay.
 During solar cycle 23, looking at protons and electrons measured by
 the 3DP experiment on board the \emph{Wind} spacecraft,
 \citet{Agueda_2011} obtained the spectrum of 1--10 MeV solar
 neutrons.  They found that the neutron spectrum escaping the Sun is
 well below the observed background proton spectrum.
 At these energies, most neutrons decay before reaching the Earth.
 Future missions such \emph{Solar Probe} and \emph{Solar Orbiter} will
 operate at radial distances $<1$ AU, where the neutron flux is higher
 (about 65\% of 10 MeV neutrons decay before reaching 0.3 AU, the
 perihelion distance of these missions) and is potentially detectable.

 The SONG instrument onboard CORONAS-F detected neutrons and
 gamma-rays in four events, on 26, 28, and 29 October and on 4
 November 2003.  The 2003-10-28 flare was the best observation, as
 reported by \citet{Kuznetsov2010}.

 More recently, \citet{Muraki_2016} reported the simultaneous
 observation of solar neutrons from the SEDA-NEM instrumend on board
 the International Space Station \cite{Koga_2011} and high mountain
 observatories located in Mt. Chacaltaya, Bolivia, and Mt. Sierra
 Negra, Mexico, associated with the flare 2014-07-08 16:06 UT.
%The measured neutron spectrum is less intense than the inferred spectrum for the flare 2015-09-07 17:34 UT from observations at Mt. Chacaltaya, represented as a power-law with spectral index $2.0\pm0.2$ between 100 MeV and 2.5 MeV. 

 Finally, a dozen flares are associated with detected neutron fluxes
 at the Earth, with space and/or ground observations.  They are
 considered in details by \citet{Yu201525}, where also the theoretical
 aspects are reviewed.

%%%%%%%%%%%%
\subsection{Galaxy center}\label{sec-gal-cen}
%%%%%%%%%%%%

 Solar flares inject neutrons in the interplanetary space with
 energies up to the GeV range and in well defined time intervals.  On
 the other extreme of the CR energy spectrum, at the so-called CR
 ankle, there is indirect evidence of neutrons coming from a region
 close to the Galaxy center with energies of order of $10^{18}$~eV.

 At these energies, measurements can only be performed by looking at
 atmospheric showers, hence it is impossible to distinguish a
 neutron-initiated cascade from a proton event.  However, the
 distribution of ultra high energy CR shows a small anisotropy which
 is not detected at lower energies.
 Such anisotropy was first reported by the AGASA air shower array at 4\%
 level around $10^{18}$~eV, and it was interpreted as an excess near
 the directions of the Galactic Center and the Cygnus region
 \cite{Hayashida_1999}.
 More recently, the Pierre Auger Observatory has also confirmed the
 presence of this anisotropy at 4.4\% level (amplitude of first
 harmonic) \cite{Aab_2015}.

 Such an excess is most easily explained in terms of neutral
 particles, which travel along the line of sight of their source.
 The Galaxy center is 7.6--8.7 kpc away, hence the range of
 $\gamma$ is (8.8--10.0)$\times 10^{8}$, which corresponds to a neutron
 energy $E =$ (8.3-9.5)$\times 10^{17}$~eV.  Hence neutrons from this region,
 if any, can only contribute to the ankle portion of the CR spectrum.
 \citet{Clay_2000} interpreted the anisotropy reported by AGASA as an
 evidence for a detection of $10^{18}$~eV neutrons, because protons
 would not travel along straight lines.

 In the CR spectrum between $10^{18}$ and $10^{20}$ eV there are two
 changes of slope, with a flux increase around $8.5 \times 10^{18}$~eV
 and a steep decrease above $4 \times 10^{19}$~eV (see e.g.~Figure
 29.8 of \cite{pdb2016}).
 The hardening of the spectrum around $8.5 \times 10^{18}$~eV is
 remarkably close to the neutron energy threshold of 8.4--8.7
 $\times10^{17}$~eV inferred from the distance of $7860\pm140$~pc of
 Sagittarius A*, where a supermassive black hole with relativistic
 jets is located \cite{Brown_1982}.
 If this were the only active source in this energy region (which is
 likely not the case), one could think that the steepening above
 $4 \times 10^{19}$~eV reflects the true neutron spectrum (e.g.~the
 spectral index and the normalization of a power-law model), because
 at these energies most neutrons reach the Earth without decaying.
 This is not in agreement with the ``classical'' explanation of the
 ankle as the region where there is a transition between a galactic
 component and an extragalactic contribution.  On the other hand, it
 is not straightforward to explain all measured details of the ultra
 high energy cosmic ray spectrum in terms of this transition (for a
 recent discussion, see e.g.\ \cite{Haungs_2015}).  Hence it is not
 excluded that a neutron component may affect the energy spectrum, in
 addition to providing a simple explanation for the detected
 anisotropy. 

 Cygnus X-1 is a black hole X-ray binary located at $2200\pm200$~pc,
 emitting gamma-rays with energies in excess of 100 GeV
 \cite{Albert_2007}.  It can contribute neutrons above $2 \times
 10^{17}$~eV, at the so-called second knee of the CR spectrum. 
 Of course, any source of high-energy neutrons is expected to emit
 also protons (and other charged particles).  However, protons diffuse
 for millions of years in the Galaxy before reaching the Earth,
 contributing to the CR spectrum with a long delay with respect to
 neutrons, which travel straight practically at the speed of light.
 Thus neutrons from Cygnus X-1 take about 6500 years to reach the
 Earth, while those from Sagittarius A* take about 25600 years, both
 arriving much before the protons ejected at the same time at the
 source can be detected.

 Although it is suggestive that the energy thresholds connected to
 Cygnus X-1 and Sagittarius A* correspond to known features of the CR
 spectrum, one should not forget that there may be additional sources,
 likely of galactic origin below the ankle and extragalactic in the
 ankle region. 

 On the other hand, also these source can emit
 neutrons, which play an important astrophysical role.
 For example, \citet{Sikora_1989} considered the creation of
 relativistic neutrons in active galactic nuclei (AGN).  They are an
 efficient way of transporting a fraction of the total energy output
 away from the AGN, thanks to their big interaction length.  Their
 decays then release this energy after 1--100 pc, producing charged
 particles (protons and electrons), which interact almost locally, and
 neutrinos.
 Furthermore, \citet{Tkaczyk_1994} concluded that about half of the
 AGN acceleration power is transferred to relativistic neutrons, which
 can then escape the region of strong magnetic field.  This makes it
 possible to transfer a large amount of mass and energy in hadronic
 jets.  The photodisintegration process of the nucleus is a powerful
 source of neutrons, because the photodisintegration time of nuclei in
 comparison to the photopion production time is an order of magnitude
 shorter, and so the nuclei are completely destroyed first.  On
 average, a nucleus with mass number $A$ generates half of its own
 mass in neutrons with the same Lorentz factor.
 % Thus the neutron luminosity can be 1.5 times greater than what is
 % expected from p-p and $\gamma$-p interactions, each source carrying
 % about one third of the initial proton luminosity.

%%%%%%%%%%%%
\subsection{Supernova remnants}\label{sec-snr}
%%%%%%%%%%%%

 Neutrons emitted during solar flares have been detected in several
 events, and there is indirect evidence for a neutron component causing
 the observed anisotropy in the ultra high energy CR spectrum.
 Supernova remnants (SNR) should also provide a detectable
 flux of neutrons, but this remained undetected so far.

 \begin{figure}
   \centering
   \includegraphics[width=\columnwidth]{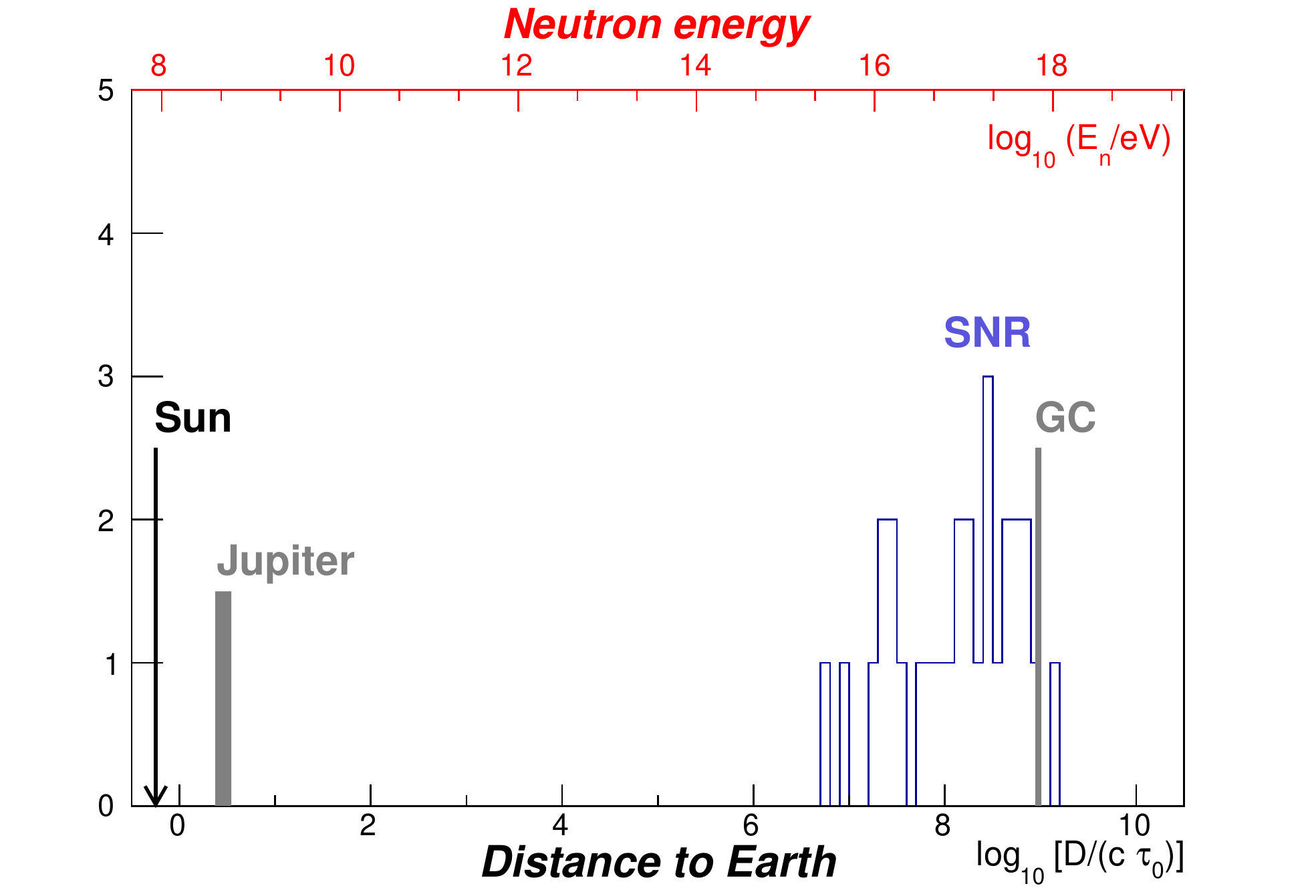}
   \caption{Distance of identified progenitors of supernova remnants,
     compared to other astrophysical objects: the Sun (at 1 AU),
     Jupiter (at variable distance of 4.2-6.2 AU from the Earth), and
     the Galactic Center (whose distance is uncertain).  The typical
     neutron energy in eV is shown by the red scale.}
   \label{fig-snr_dist}
 \end{figure}

 The SNR distances range from 150 ly to 39000 ly
 \cite{SIMBAD}\footnote{Data from \cite{SIMBAD} summarized in
   \url{https://en.wikipedia.org/wiki/List_of_supernova_candidates}.},
 i.e.\ from $5.3 \times 10^6 \,c\tau_0$ to
 $1.4 \times 10^9 \,c\tau_0$.  Hence the energy threshold varies from
 $5.0 \times 10^{15}$ eV to $1.3 \times 10^{18}$ eV, i.e.\ it spans
 the region of the CR spectrum encompassing the knee and the second
 knee, where the measured CR spectrum (with air shower detectors)
 shows a few changes in slope (see Figure 29.8 of \cite{pdb2016}).
 The distribution of the logarithm of the SNR distance
 (figure~\ref{fig-snr_dist}) is slightly asymmetric, with a broad peak
 close to 8.5, which corresponds to an energy scale of about
 $3 \times 10^{17}$ eV.  The median is 8.24, corresponding to
 $E = 1.6 \times 10^{17}$ eV.  At these energies, the CR spectrum has
 a feature called second knee.  The 0.25-quantile corresponds to
 $3.7 \times 10^{17}$ eV, just above the slope change between the knee
 and the second knee.  Finally, the 0.75-quantile corresponds to
 $4.0 \times 10^{17}$ eV, a region in which a single power-law seems
 to correctly model the CR energy spectrum up to the ankle.

 Together with the correspondence between the ankle and the threshold
 energy for neutrons coming from the Galaxy center, the energy scales
 related to the SNR distances are suggestive of a non coincidental
 connection with the shape of the CR spectrum at very high energies.
 Apart from the long time delay between the arrival of neutral and
 charged particles to the Earth, the problem is that in this energy
 region there is no way of identifying the parent particle, which
 gives rise to the hadronic cascade in the Earth atmosphere.
 The author is not aware of any simulation, in which ions and neutrons
 are emitted from several point sources corresponding to the known
 supernova remnants, and are traced inside the Galaxy until the reach
 the Earth.
 However, the absence of a significant anisotropy below the ankle
 implies that free neutrons cannot be a large fraction of the detected
 particles.

%%%%%%%%%%%%
\subsection{Cosmic neutrons}\label{sec-cosmic}
%%%%%%%%%%%%

 In the wide energy range from solar flares to supernova remnants,
 there is no known point source of neutrons, which can reach the Earth
 before decaying.  Thus, one is left with the diffuse background of
 secondary cosmic neutrons produced by the interactions of cosmic rays
 with the interstellar medium and with those produced in the stellar
 and planetary atmospheres.
 The flux of the diffuse component of cosmic neutrons was never
 detected so far, and at low energies is much smaller than the albedo
 component due to CR initiated hadronic showers in the Earth
 atmosphere (discussed below in section~\ref{sec-albedo}).

 As an order of magnitude estimate for the diffuse cosmic neutron
 flux, one can take the CR antiproton flux, recently measured by
 PAMELA \cite{Adriani_2010} and AMS-02 \cite{AMS2antiprotons} from
 60~MeV up to few hundred GeV.  The simplest process with production
 of antiprotons is the collision of a CR proton with a hydrogen atom
 of the interstellar medium:
 $\text{p} + \text{p} \to \text{p} + \text{p} + \text{p} +
 \bar{\text{p}}$.  The energy threshold (1.88 GeV in the center of
 momentum system) is higher than the threshold (1.08 GeV) for the
 simpler reaction with a secondary neutron in the final state,
 $\text{p} + \text{p} \to \text{p} + \text{n} + \pi^+$.  This means
 that the low-energy region is a bit harder for the neutrons than for
 the antiprotons, although it does not make any sizable difference at
 high energies.

 The number density of secondary antiprotons provides a lower bound on
 secondary neutrons, because free neutrons can also be produced by
 spallation, as fragments of larger nuclei which act as targets for CR
 protons.  However, spallation is not expected to inflate
 significantly the number density of free neutrons, because elements
 heavier than hydrogen are present only in small amounts: the number
 density of He is only about 10\% of H, and the sum of O, Ne, N and C (the
 next most abundant species in the universe) is about 0.1\% of the
 hydrogen density.
 % In addition, it is likely that fragments are produced in the form of
 % alpha particles, which have large binding energy.  Even in a
 % collision p-He it is likely that the fragments are deuterons rather
 % than neutrons.

 On the other hand, the flux of charged particles can be significantly
 different from the flux of neutral particles with the same number
 density.  The reason is magnetic deflection: while a neutral particle
 travels along a straight path and has only one chance to cross a
 detector, a charged particle may be deflected by the magnetic field
 and cross more than once a given surface.  This effect is evident for
 the trapped protons and electrons in the Earth magnetosphere (an
 effect known since decades, and measured with unprecedented precision
 by AMS-01 \cite{2000PhLB..472..215A,2000PhLB..484...10A} and PAMELA
 \cite{2015ApJ...799L...4A,Mikhailov_2016,Bruno_2016}), and should
 work similarly over wide regions if the magnetic field intensity is
 smaller.  Thus, the measured flux of charged particle is enhanced by
 any sort of trapping, whereas the flux of neutral particles is
 independent of magnetic fields.  However, at present is not known by
 how much the charged particle flux in the solar system is enhanced.

 One further complication with neutrons is that they may be also
 created by electron capture from protons and electrons trapped in the
 same region.  This is suggested by the comparison between the neutron
 measurements in the Van Allen belts by the Russian missions
 \emph{Salut-7} and \emph{Kosmos-1686} \cite{Bogomolov_1998} and the
 earlier measurements by COMPTEL \cite{Morris_1995} and several
 balloon flights (more details in section~\ref{sec-albedo}), which all
 agree on neutron fluxes smaller by one order of magnitude.

 If the size of the trapping region (which has to be not too far from
 the solar system) is not much bigger than $\gamma c \tau_0$, then the
 flux of neutrons with Lorentz factor $\gamma$ is enhanced, where
 $\gamma$ is of the same order of magnitude of the proton and electron
 Lorentz factors.  On the other hand, the larger is $\gamma$ the
 larger is the rigidity $R$ of the parent charged particles, hence the
 larger is the curvature radius $r \simeq R/(Bc)$ for a given magnetic
 field intensity $B$.  For a proton with rigidity of 1 GV in a region
 with field intensity of 1 \micro{G}, the curvature radius is
 $10^{-5}$~pc, of the same order of magnitude as $c\tau_0$.  The same
 field will trap protons and electrons of 10 GV in a region with
 linear size comparable to $10c\tau_0$, where neutrons with
 $\gamma\approx10$ may be created by electron capture.  Hence in the
 simplest model in which the neutron Lorentz factor is the same as
 that of the parent proton and electron, one has neutronization
 happening over distances which are of the order of the neutron decay
 length at all momenta below some critical limit, above which the
 number of protons and electrons escaping from the region becomes
 important.
 So below this limit, the enhancement of the secondary neutron flux is
 almost independent of $\gamma$.

 % r = 10 pc (E/PeV) / (B/µG)
 %   = 1e-5 pc (E/GeV) / (B/µG)

 In absence of a detailed simulation of the neutral and charged CR
 flux in our galactic environment, we take the CR antiproton flux as a
 first estimate of the neutron flux. 
 The antiproton flux measured by PAMELA increases from about 
 $7 \times 10^{-3} \um{(m}^2 \um{sr} \um{s} \um{GeV})^{-1}$ between 50
 and 100 MeV, to
 $27 \times 10^{-3} \um{(m}^2 \um{sr} \um{s} \um{GeV})^{-1}$ between
 2.6 and 3.0 GeV, then decreases at the level of
 $5 \times 10^{-3} \um{(m}^2 \um{sr} \um{s} \um{GeV})^{-1}$ at 10 GeV,
 $7 \times 10^{-4} \um{(m}^2 \um{sr} \um{s} \um{GeV})^{-1}$ at 20 GeV,
 down to $2 \times 10^{-5} \um{(m}^2 \um{sr} \um{s} \um{GeV})^{-1}$
 above 50 GeV \cite{Adriani_2010}.
 The measurement by AMS-02 gives an antiproton flux reaching a maximum
 of about $1.7 \times 10^{-2} \um{(m}^2 \um{sr} \um{s} \um{GeV})^{-1}$
 at 3 GeV, consistent with the PAMELA result within the uncertainties,
 decreasing by a factor of 10 at 16 GeV, % 1.7e-3
 by another factor of 10 at 35 GeV, % 1.7e-4
 by another factor of 10 at 80 GeV, % 1.7e-5
 by another factor of 10 at 200 GeV, % 1.7e-6
 until reaching
 $2 \times 10^{-7} \um{(m}^2 \um{sr} \um{s} \um{GeV})^{-1}$ between
 260 and 450 GeV \cite{AMS2antiprotons}.

 At rigidities high enough that the solar modulation can be neglected,
 the antiproton to proton ratio is almost constant, just below
 $2 \times 10^{-4}$.
 This means that a neutron detector in space must achieve a
 separation power of at least $10^{-5}$ between neutrons and protons,
 which should be feasible because the latter can be very efficiently
 vetoed by surrounding the detector with an anticoincidence
 scintillating layer.

%%%%%%%%%%%%
\subsection{Atmospheric neutrons}\label{sec-albedo}
%%%%%%%%%%%%

 A detectable flux of secondary neutrons is produced by CR-induced
 hadronic showers in the Earth atmosphere, which provide a steady
 state source with slow time variations connected to the solar
 activity. 
 Secondary neutrons may be detected promptly or after they have lost
 most of their initial energy in elastic collisions with the air
 molecules.  At low energy, thermalized neutrons can be captured by
 atoms, which de-excite by emitting gamma-rays.  The absorption
 $(\text{n},\gamma)$ process contributes up to few tens keV, as
 noticed by \citet{Hess_1959}, who measured the neutron spectrum from
 0.01 eV to 10 GeV in the Earth atmosphere with a series of balloon
 experiments in the fifties.
 Below 10 keV the neutron flux scales roughly as $1/E$, as it would be
 expected in an infinite nonabsorbing medium in which neutrons slow
 down by collisions with an energy-independent cross-section.
% \citet{Hess_1959}
 They found that the main deviation from $1/E$ arises from the energy
 dependence of the cross section, which flattens a bit the spectrum,
 giving a flux proportional to $E^{-0.88}$ from about 1 eV to 50 keV.
 Below 1 eV neutron absorption by nitrogen is an important process,
 competing with thermalization, and the flux is damped.
 At 1 MeV the shape is affected by the source contribution of
 ``nuclear evaporation'' % (fission)
 process, producing a bump with a very broad peak at about 500 keV.
 Beyond this bump, the spectrum falls down approximately as $1/E$,
 becoming softer above $\sim100$ MeV.

 \citet{Eyles_1972} measured the neutron ``albedo'' between 50 and 350
 MeV with balloon flights between 1967 and 1970, and found no evidence
 for a primary flux of neutrons.
 Other measurements have been performed more or less in the same
 period e.g.~by \citet{Preszler_1974} and \citet{Kanbach_1974},
 confirmed later on by the COMPTEL gamma-ray instrument on board of
 \emph{CGRO} \cite{Morris_1995}.

 Using data from the Russian missions \emph{Salut-7} and
 \emph{Kosmos-1686}, \citet{Bogomolov_1998} measured the neutron flux
 under the Earth radiation belts with a rigidity cutoff of 4.5 GV,
 finding a spectrum with shape similar as (but not identical to) the
 previous measurements, but almost one order of magnitude more intense
 than the measurements reported above.
 Unless this is due to some systematic uncertainty in the absolute
 calibration, this result would imply that the electron capture
 reaction from the protons and electrons trapped in the Van Allen
 belts enhances the flux of neutrons.

 Neutron flux measurements have been also performed in order to
 estimate the radiation effects on airline crews and atronauts. 
 In particular, \citet{Badhwar_2001} measured the 1--14 MeV neutrons
 onboard the NASA space shuttle during the STS-31 mission,
% , finding a neutron rate at 1
%  MeV of about 1.5 Hz/cm$^2$, to provide a figure of merit.
%
 whereas \citet{Matsumoto_2001} performed a measurement during the
 STS-89 space shuttle mission from thermal energies up to 100 MeV,
 confirming that the neutron flux depends on the amount of trapped
 low-energy particles.  The measured neutron rates in the equatorial
 region (about 0.2 Hz/cm$^2$ at 1 MeV) are about 5 times smaller than
 in the polar region (1 Hz/cm$^2$ at 1 MeV), which in turn is smaller
 than in the South Atlantic Anomaly (5 Hz/cm$^2$ at 1 MeV).
 Later, \citet{Koshiishi_2007} published the orbit-averaged neutron
 spectra up to 100 MeV, obtained on the International Space Station
 while orbiting in different geomagnetic regions from March to
 November 2001.  The shape is similar to what is reported by
 \citet{Matsumoto_2001} and the normalization is in between the
 equatorial and polar regions, as expected.

\begin{figure}
   \centering
   \includegraphics[width=\columnwidth]{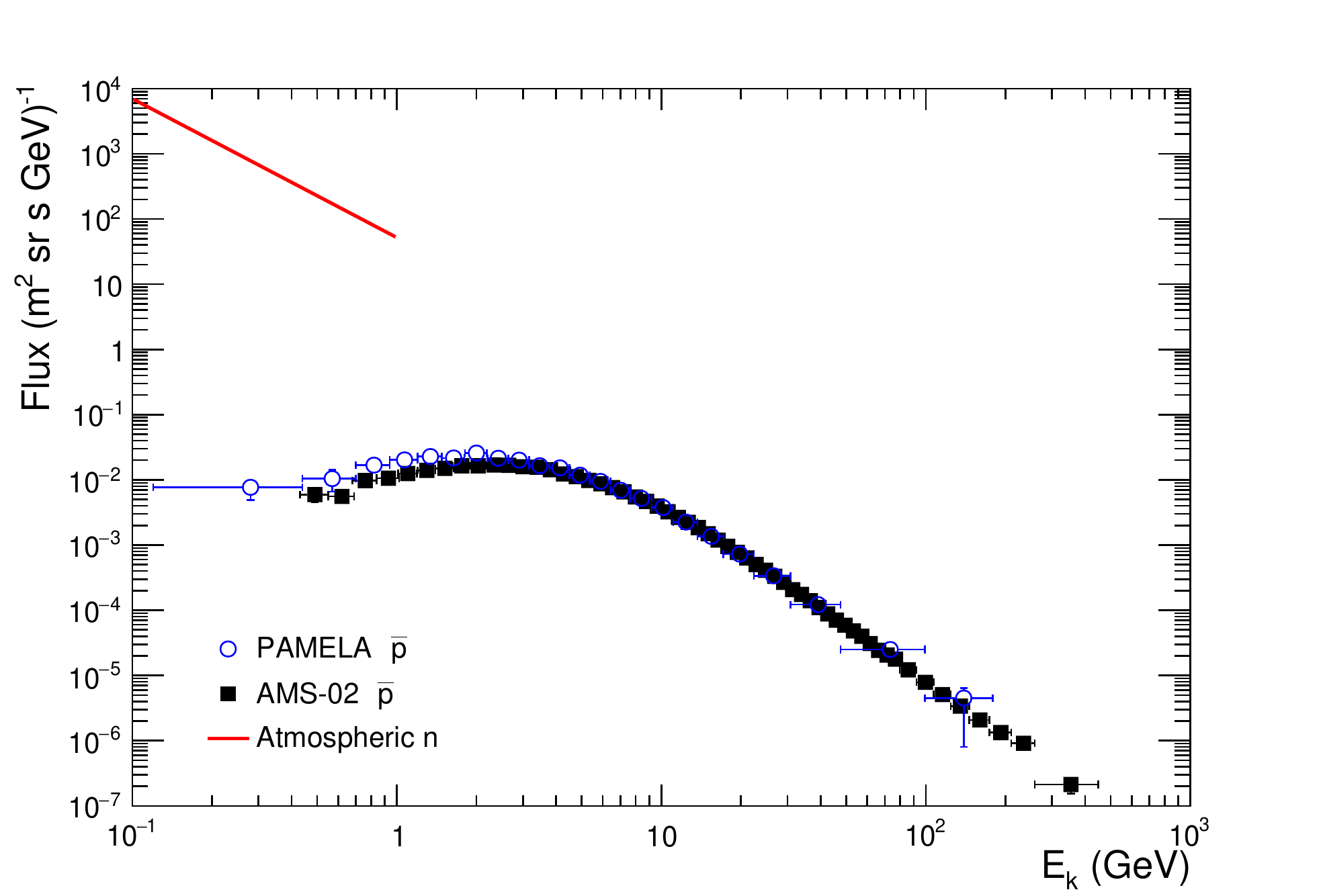}
   \caption{Comparison between albedo neutrons (continuous line) and
     antiproton flux.  The flux of atmospheric neutrons is a power-law
     with spectral index 2.13 \cite{Morris_1998}.  Data points
     represent PAMELA \cite{Adriani_2010} and AMS-02
     \cite{AMS2antiprotons} antiproton measurements.}
   \label{fig-neutrons}
 \end{figure}

 In order to provide an estimate of the neutron flux at the top of the
 atmosphere, we quote here few values:
 about 1 kHz/(m$^2$ MeV) at 1 MeV and %\cite{Hess_1959,Armstrong_1973},
 0.1 kHz/(m$^2$ MeV) at 10 MeV \cite{Hess_1959,Armstrong_1973},
 20--30 Hz/(m$^2$ MeV) at 50 MeV \cite{Hess_1959,Armstrong_1973,Morris_1998},
 about 8 Hz/(m$^2$ MeV) at 100 MeV and %\cite{Morris_1998},
 0.7 Hz/(m$^2$ MeV) at 300 MeV \cite{Morris_1998}.
 The rate becomes about 10 times higher inside the geomagnetic belts,
 where the soft trapped protons and electrons may give birth to
 neutrons via electron capture.
 In the same energy region, from time to time neutrons from solar
 flares dominate the rate for a relatively short duration.
 Anyway, even in quite Sun periods at low energy --- measurements are
 available only up to few hundred MeV --- the secondary neutrons
 produced in the Earth atmosphere are many more than the expected
 diffuse cosmic neutron flux (figure~\ref{fig-neutrons}), hence it is
 fundamental to be able to separate these two components.

 For a detector orbiting outside the Earth atmosphere, secondary
 neutrons mostly comes from below (hence the name ``albedo
 neutrons''), with a flux which is expected to be sizable up to the
 tangential direction and decrease significantly when looking upward.
 This means that the background from albedo neutrons can be suppressed
 using a directional detector, or by performing a differential
 measurement comparing zenith and nadir fluxes.

 The Earth magnetic field affects the flux of incoming charged
 particles, reflecting low-energy particles as a function of their
 magnetic rigidity.
 Hence the neutron spectrum was studied by \citet{Morris_1995} as a
 function of the geomagnetic cutoff and of the geocentric zenith
 angle.  The measurement is well fitted by the product of an
 exponential term in cut-off rigidity, a linear function of the zenith
 angle (ranging from 3$^\circ$ to 177$^\circ$), and a factor
 accounting for the solar modulation.  The linear dependence on the
 zenith angle is attributed to the screening of the \emph{CGRO}
 spacecraft mass, which is much more important than the production of
 secondary neutrons in the spacecraft itself.
%
 % \citet{Morris_1995} show the neutron flux with the instrument
 % pointing down and rigidity cut-offs of 4.5 GV and 8.5 GV.  The former
 % spectrum is well modeled by a broken power-law, and is consistent
 % with the measurements by \citet{Kanbach_1974} and
 % \citet{Preszler_1976} and is similar to the spectrum computed by
 % \citet{Armstrong_1973}, who calculated the neutron flux spectra
 % induced by galactic CR protons and alpha particles in the Earth
 % atmosphere at solar minimum and 42$^\circ$ N ($E_c=3.8$ GeV for
 % protons and 6.3 GeV for He nuclei) from $10^{-8}$ to $10^5$ MeV at
 % different depths.  On the other hand, for 8.5 GV cutoff the spectrum
 % is harder and the flux above 10 keV is smaller (as expected).
%
 Later, \citet{Morris_1998} reanalyzed the same COMPTEL data and
 fitted them together with previous measurements, obtaining a broken
 power-law with spectral indices of 0.56 below 67 MeV and 2.13 above
 this energy.  The normalization is found to depend on the Mount
 Washington neutron monitor rate.
 % (an explicit formula is provided,
 % expressing the neutron flux as a function of the zenith angle, the
 % rigidity cutoff, the neutron energy, and the Mt. Washington rate).
 % When the latter has value 2340, the neutron flux at 100 MeV is about
 % $10^{-3}$ Hz/(cm$^2$ MeV).

%%%%%%%%%%%%%%%%%%%%%%%%%%%%%%%%%%%%%%%%%%%%%%%%%%%%%%%%%
\section{Dark matter searches}\label{sec-DM}
%%%%%%%%%%%%%%%%%%%%%%%%%%%%%%%%%%%%%%%%%%%%%%%%%%%%%%%%%

 As we have seen, solar flares are the only known astrophysical source
 of neutrons, which have been detected so far.  Solar neutrons may
 reach the Earth from about 20 MeV (below which they mostly decay in
 flight \cite{Frye_1988}) to few GeV (above which the injected flux
 becomes negligible), with a tight time correlation with X-ray flares.
 Hence solar flare neutrons are easily identifiable events.

 If an instrument in orbit is able to recognize secondary neutrons
 produced by CR interactions in the Earth atmosphere, what remains is
 a very low cosmic neutron flux, which so far remained undetected,
 spanning a wide energy range up to the CR knee region, above which
 supernova remnants are expected to contribute (however, at these
 energies one cannot identify the incoming particle and the only
 possibility is to perform statistical studies of the hadronic showers
 produced in the atmosphere).

 Thus, the diffuse cosmic neutron flux is a (very low) background for
 searches for exotic signatures, like the final products of the
 annihilation of dark matter (DM) particles.  Hence it is interesting
 to know which energy regions are best suitable for DM searches with a
 neutron detector in space.

 The chapter on DM searches of the latest Review of Particle Physics
 \cite{pdb2016} lists a number of candidates, like primordial black
 holes, axions, sterile neutrinos, and weakly interacting massive
 particles (WIMPs).  Among them, primordial black holes are not
 directly detectable, although their accretion jets may well emit
 neutrons with a wide energy range, as we mentioned about AGNs in
 section~\ref{sec-gal-cen}.  Thus they are not considered in this
 section.  The mass range for axions is within the thermal range for
 neutrons, at which neutron capture takes place.  The albedo component
 is overwhelming in this case, because there is no directional
 detection, hence axions are not considered in the following.  Sterile
 neutrinos with keV masses are invisible for an orbital detector: even
 in case one could detect a nuclear recoil from such a low energy
 particles, the sensitive volume would need to be enormous because of
 the extremely small neutrino cross-section. 
 So we are left with WIMPs, which are the most promising candidates
 for the ``cold DM models'' which best fit all available astrophysical
 and cosmological data (see \cite{pdb2016} and references therein).
 WIMPs masses may range from about 10 GeV to few TeV, which is the
 energy range in which the cosmic neutron background is very low.
 Lighter candidates have been also proposed, like strongly interacting
 massive particles (SIMPs) with masses in the MeV to GeV range
 \cite{Hochberg2014}.  They may also annihilate in fermion-antifermion
 pairs, like WIMPs.

 The current approach for indirect detection of DM particles with
 masses from MeV to TeV is to look for a ``bump'' in the cosmic gamma,
 positron and antiproton fluxes, as currently being done by different
 space experiments.  However, the same decays that create
 antiproton-proton pairs will also create antineutron-neutron pairs.
 Hence, one may also measure the cosmic neutron flux and look for an
 excess in some energy range.  The advantage of this approach is that
 neutrons provide directional information.  For example, massive
 objects like the Sun or Jupiter may attract DM particles increasing
 their local density, and so also their interaction rate. 
 Thus, an excess of neutrons coming from the Sun in periods without
 solar flares could be interpreted in terms of local DM density.

 Note that \cite{Salucci2010} obtained a DM density of $0.43\pm0.15$
 GeV/cm$^3$ in the galactic region of the solar system, a bit higher
 than the galactic average commonly used in the literature (0.3
 GeV/cm$^3$).   However this does not account for the presence of
 compact massive objects like the Sun.  Indeed, the gravitational
 field of the Sun may locally enhance the DM density by orders of
 magnitude.  However, any uncertainty on the galactic DM density is
 also amplified the same way, making an estimate of the density around
 the Earth unfeasible
 \footnote{
   A computation carried on in the hypothesis of an ideal gas at
   constant temperature in a central classical gravitational field
   gives a radial density function proportional to $\exp(A/r)/r^2$,
   where $A = G M M_{\odot}/(k_B\,T)$ is a constant length.  The
   temperature of the cosmic background radiation is 2.7~K and the
   that of the cosmic neutrino background is 2~K.  The DM temperature
   shall be lower.  At $T=1$~K one gets $A=5.55$~pc, while at 0.1~K
   one gets 55.5~pc, to make an example.
     % In the Big Bang scenario, particles that interact with the
     % primordial plasma only through the weak interaction
     % (e.g. neutrinos) decouple around 1 MeV.  However DM particles are
     % expected to decouple somewhat earlier, at any temperature greater
     % than 1 MeV and (likely) smaller than QCD phase transition (150
     % MeV).  The decoupling temperature for WIMPs is not known.
   Hence one may get a huge amplification of the DM density close to a
   massive object.  The problem is that, due to the exponential
   dependence, any small uncertainty on the local interstellar density
   maps into a huge uncertainty on the amplified DM density at 1 AU
   from the Sun.  This means that it's almost impossible to estimate
   the DM at the Earth from the DM density obtained with measurements
   of galactic parameters.}.

 Among the direct detection attempts \cite{Mayet20161}, there is a
 positive measurement claimed by the DAMA/LIBRA collaboration
 \cite{Bernabei2008, Bernabei2010}, who found an annual modulation
 with the expected period (1 year) and phase (maximum around June 2)
 for a distribution of DM particles crossed by the solar system, with
 the Earth moving forward or backward with respect to the Sun motion
 every 6 months.  Interpreted in terms of WIMPs, two scenarios appear
 most plausible: a 50 GeV mass with cross-section of
 $7 \times 10^{-6} \um{pb}$ or a particle with mass of 6--10 GeV and
 cross-section of order $10^{-3} \um{pb}$ \cite{pdb2016}.  However,
 this result was criticized and could not be reproduced by other
 direct-detection experiments, which instead rejected the signal
 claimed by DAMA/LIBRA.  Nevertheless, the annual modulation detected
 by DAMA
 % (7 years exposure with 250 kg active material)
 and LIBRA
 % (6 years exposure and 100 kg of detectors)
 could not be explained by some other mechanism so far.  Hence these
 mass regions remain interesting.

 Other (debated) positive results have been also claimed by other
 experiments.
 CRESST found a signal compatible with WIMP masses of about 11 GeV or
 22 GeV \cite{Angloher2012}.
 CDMS found a signal compatible with a WIMP mass of about 10 GeV
 \cite{PhysRevLett.111.251301}. 
 CoGeNT also found some evidence for an annual modulation compatible
 with a WIMP mass of 7 GeV \cite{PhysRevLett.107.141301}.

 Note that radio observations of the Ursa Major II dwarf spheroidal
 galaxy \cite{PhysRevD.88.083535}, interpreted in terms of possible DM
 annihilation into e$^+$e$^-$ producing synchrotron radiation in the
 magnetic field of the dwarf galaxy, exclude 10 GeV WIMP for
 $B>0.6$~\micro{G} at the thermal rate
 $\langle\sigma v\rangle = 2.18 \times 10^{-26}$ cm$^3$/s at $2\sigma$
 level.  However, they do not constrain decays into other fermion pairs.

 On the side of indirect detection, interesting results emerge both
 from gamma-ray and from cosmic-ray measurements.
 \emph{Fermi}-LAT data show an excess of GeV photons from an extended
 region around the Galaxy center \cite{Hooper2011412,Daylan20161}.
 Explanations in terms of standard objects, like a population of
 thousands unresolved millisecond pulsars, might be able to explain
 the emission from the Galactic Center, but are challenged by the fact
 that the signal extends well beyond the boundaries of the central
 stellar cluster \cite{Daylan20161}.
 On the other hand, models with DM particles of 7--10 GeV, which
 annihilate primarily into a $\tau^+ \tau^-$ final state but also into
 hadronic final states, well reproduce the data.  However,
 \citet{Daylan20161} emphasize that better fits are obtained with
 heavier particles (with masses in the range 20--60 GeV) annihilating
 mostly into quarks.
 These mass ranges are remarkably similar to those preferred by DAMA/LIBRA.
% However \citet{PhysRevD.83.035022} argues that 

 Another interesting result is the excess of photons around 130 GeV
 discovered in 2012 using public \emph{Fermi}-LAT data
 \cite{Weniger_2012} but no longer visible in a reanalysis of the same
 data \cite{PhysRevD.91.122002} and also excluded by H.E.S.S.\ 
 \cite{PhysRevLett.117.151302}. 
 Interestingly, WMAP data have been also interpreted as evidence for
 annihilation of WIMPs with masses above 100 GeV
 \cite{PhysRevD.76.083012}, however there is no consensus on this
 \cite{Dobler_2012}.

 Finally, the CR electron and positron measurements also reveal
 interesting details.  The positron fraction measured by PAMELA
 \cite{PhysRevLett.111.081102} and AMS-02
 \cite{PhysRevLett.113.121101} increases between 10 and 200 GeV, and
 then flattens above 200 GeV.  The behavior of the positron fraction,
 considered together with the rather hard electron spectrum measured
 by PAMELA \cite{PhysRevLett.106.201101}, AMS-02
 \cite{PhysRevLett.113.121102} ATIC \cite{ATICnature07477}, Fermi-LAT
 \cite{PhysRevLett.102.181101} and H.E.S.S. \cite{Aharonian_2009}
 between 100 and 1000 GeV, can in principle be explained through WIMP
 annihilation.  However, it is likely that nearby and recend supernova
 or pulsar sources of electrons and positrons can also explain the
 observed features.
 % This would disfavor WIMP masses below 300 GeV annihilating directly
 % into leptons \NdA{PhysRevD.83.035022}.
%
 % \citet{PhysRevLett.117.091103} summarize the latest AMS-02
 % measurements by stating that, from about 60 to 500 GV, the
 % antiproton, proton, and positron fluxes have nearly identical
 % rigidity dependence, different from the electron flux.

 WIMP annihilations would also produce detectable features in the
 antiproton and antideuteron spectra \cite{Aramaki20161}.  Using the
 latest AMS-02 antiproton measurement, two independent groups
 \cite{PhysRevLett.118.191101,PhysRevLett.118.191102} claimed that a
 dark matter signal with mass between 20 and 80 GeV and
 velocity-averaged hadronic annihilation cross-section of order
 $10^{-26} \um{cm}^3 \um{s}^{-1}$ is able to reproduce the data much
 better than the background-only hypothesis.  While the analysis in
 \cite{PhysRevLett.118.191102} has a net preference for masses of
 about 70--80 GeV and
 $\langle\sigma v\rangle \approx 3 \times 10^{-26}$ cm$^3$/s,
 \cite{PhysRevLett.118.191101} obtain a more diagonal contour map
 extendind to lower masses and velocity-averaged cross-sections, with
 a more pronounced dependence on the parametrization of the antiproton
 production cross-section.

 This suggests that some signature should also be visible in the
 neutron spectrum, at the same energies.  However, neutrons have
 shorter range than electrons and positrons, hence probe a small
 portion of the Galaxy around the solar system.  For example, 80 GeV
 neutrons come mostly from a sphere of radius 142~AU (a bit larger
 than the solar system), while positrons come from distances bigger
 than 1 kp (of order of the thickness of the galactic disk)
 \cite{KOBAYASHI2001653}.  Both sample a much smaller region than
 protons and He nuclei, diffusing for tens of millions years in a
 large fraction of the Galaxy.

 A lower limit to the DM contribution to the local neutron flux can be
 obtained from the galactic averaged DM density.  For example, 80 GeV
 neutrons might come from the annihilation of $\sim80$ GeV WIMPs, whose
 galactic-averaged number density $n_{80}$ would be about 0.005~cm$^{-3}$.
 The annihlation rate would be
 $\Gamma_{80} = \frac{1}{2} n_{80}^2 \langle\sigma v\rangle$ which, using
 $\langle\sigma v\rangle = 3 \times 10^{-26}$ cm$^3$/s,
 gives $\Gamma_{80} = 3.8 \times 10^{-31} \um{cm}^{-3} \um{s}^{-1}$.
 We are only interested into annihilations happening inside a sphere
 with radius $\gamma c\tau_0 = 2.1 \times 10^{15} \um{cm}$, that is
 inside a volume $V = 4 \times 10^{46} \um{cm}^3$.
 The annihilation rate inside this volume is
 $\Gamma V = 1.5 \times 10^{16} \um{s}^{-1}$ and, to first
 approximation, this is also the production rate of neutrons.
 As neutrons decay in the lab-frame after
 $\gamma\tau_0 = 7 \times 10^{4} \um{s}$, DM annihilations provide a
 stationary neutron population inside $V$ with density
 $N/V = \Gamma \gamma\tau_0 = 2.5 \times 10^{-26} \um{cm}^{-3}$. 
%
% $\Gamma V \gamma\tau_0 = 1 \times 10^{21}$ neutrons inside V
%
 These neutrons have a homogeneous and isotropic distibution of
 particles traveling at the speed of light. 
 A given flat surface $A$ (say the entrance face of a detector) is
 reachable by all neutrons contained in a hemisphere centered on $A$
 with radius $\gamma c\tau_0$. The rate at which they cross $A$ is
 $r = 2\pi \gamma c A N/V$.  If $A = 1 \um{m}^2$,
 $r = 3.8 \times 10^{-10} \um{s}^{-1}$, which is extremely low.
 This lower limit applies in the interstellar space, far from the
 gravitational influence of star-like bodies.  However around the
 Earth the DM density should be much higher than the galactic average
 because of the Sun gravitational field.  Hence it is well possible
 that the local rate is several orders of magnitude bigger \cite{Note2}.

 In addition to the amplification of the DM signal close to star-like
 objects mentioned above, there is another interesting mechanism that
 can increase the neutron flux from DM, not by incrementing the DM
 density but by extending enormously the sampled volume.  This is the
 DM ``transporting'' mechanism, which \cite{Kim2017} invoke to
 reconcile the positron measurements with antiproton data.  As
 mentioned above, positrons and electrons probe a volume considerably
 smaller than protons and antiprotons.  Hence positron data,
 interpreted in terms of a DM source, imply a too large cross-section
 or density compared to computations performed with average galactic
 values, which reproduce antiproton data.  However, the two types of
 measurements can be reconciled in case DM annihilations, happening
 mostly in the central region of the Galaxy, produce an unstable
 intermediate dark-sector state, which subsequently decays into a
 lighter DM particle in the vicinity of the Earth.

 Although the specific model presented in \cite{Kim2017} does not
 enhance the local neutron flux, because it is based on a dark-sector
 state which pairs preferentially to leptons, the authors confirm that
 the same theoretical approach can work as well if light quarks are
 produced in the decay of the intermediate state.  In particular, the
 accessible volume increases proportionally to the decay range of the
 dark-sector particle, which depends on the mass difference between
 the heavier DM particle and the invisible carrier.  As the accessible
 volume scales as the third power of the Lorentz factor, an invisible
 carrier which is significantly lighter than the progenitor DM
 particle extends the volume contributing to the local flux of
 neutrons by few orders of magnitude.
 In this model low-energy neutrons may derive from DM annihilations
 in distant regions, because the actual neutron production happens
 close to the detector.  In addition, one may get relatively soft
 neutrons (say below 10 GeV) also from a heavy DM particle (say above
 100 GeV), depending on the mass of the intermediate invisible state
 and its decay.

 In summary, neutrons are created by the same reactions that produce
 antiprotons, including possible WIMP annihilations or decays.
 Contrary to antiprotons and positrons, neutrons provide directional
 information, which is unavailable with charged particles.  Although
 neutrons may come from a limited volume, in the energy range which is
 relevant for DM searches there is no other ``background'' source of
 neutrons.  Thus, any detection of cosmic neutrons with a precise
 energy signature would be a smoking gun pointing toward the existence
 of new particles.

%%%%%%%%%%%%%%%%%%%%%%%%%%%%%%%%%%%%%%%%%%%%%%%%%%%%%%%%%
\section{Neutron detection}\label{sec-detection}
%%%%%%%%%%%%%%%%%%%%%%%%%%%%%%%%%%%%%%%%%%%%%%%%%%%%%%%%%

 The most interesting mass range for WIMPs goes from few GeV to
 few hundred GeV.  Hence the ideal neutron detector should have
 the following characteristics.  First, it should operate in an energy
 range which covers the solar flare emissions at the low-energy side,
 and reach the highest possible range (ideally up to the TeV
 scale, but this is hardly possible).
 Second, it should possess directional capability, in order to locate
 the source of each neutron.  Third, it should operate in space, such
 that the atmospheric neutrons have a well defined directional
 distribution, and can be separated from the other sources, including
 the diffuse cosmic neutron background.  Fourth, it should have an
 excellent rejection power for charged cosmic rays, and allow for
 pulse-shape discrimination to efficiently separate gamma rays from
 neutrons.  Finally, it should have sufficiently good energy
 resolution to provide an estimate of the WIMP mass in case of the
 detection of an excess over the background.

 Of course, achieving all this goals at the same time is quite
 difficult, if not impossible.  The main challenge is represented by
 the trade between the energy range, the detector size, and the
 directionality.  Although it is desirable to reach very high energies
 in order to probe the WIMP parameter space up to few hundred GeV
 mass, the expected neutron flux is a falling function of the energy
 and the counting rate becomes quickly negligible unless the effective
 area is very large.
 A $4\pi$ geometry has the largest effective area, but it is difficult
 to reconcile with the need of reconstructing the incoming neutron
 direction.  For example, \citet{pioch2011} showed that a Bonner
 sphere spectrometer provides sensitivity up to 100 GeV, but this kind
 of detector does not provide information about the incoming
 direction, because neutrons must be thermalized before being
 detected.
 A directional neutron detector with large effective area, which is
 able to measure high energies, must necessarily be a large-volume
 heavy-material instrument.  As such, it is very hard to put it in
 orbit.  Whereas at low energies (up to few GeV) the n-p elastic
 scattering can be exploited to reconstruct the initial neutron
 momentum (see below), at high energies inelastic scattering dominates
 and hadronic showers are produced.  By reconstructing the shape of
 the shower, one may find the initial direction and the total neutron
 energy, but hadronic cascades develop over wide volumes and total
 containment requires a very large detector.  At the same time, one
 should choose an active material which possesses heavy nuclei, to
 increase the interaction probability, which again goes toward larger
 detector masses.  The alternative is to build a sampling calorimeter,
 in which the active material can be plastic scintillator, but the
 final result is still very massive, because of the required layers of
 heavy material that are needed to improve the development of a
 hadronic cascade.

%%%%%%%%%%%%
\subsection{Kinematics of elastic scattering}\label{sec-elastic}
%%%%%%%%%%%%

 Assume that the neutron undergoes two interactions in the detector
 volume, and that the first is an elastic interaction in which a
 hydrogen nucleus (considered as a free proton) is kicked out of a
 molecule.  The best active medium for the first scattering is thus an
 organic scintillator.
 Let $n,n',p$ represent the 4-momenta of the neutron before and after
 the first collision, and the 4-momentum of the proton emitten in the
 first elastic scattering, respectively.
 Next, assume that the scattered neutron interacts a second time
 inside the detector.  
 % The second collision may be another elastic scattering or an
 % inelastic interaction in a heavy material.  The latter is more likely
 % to happen in a heavy medium like BGO, which can be used to build up a
 % hadronic calorimeter.
%
 Finally, assume that the detector is able to measure the
 positions A and B of the first and second neutron collisions and the
 4-momentum of the proton scattered at point A.

 \begin{figure}
   \centering
   \includegraphics[width=\columnwidth]{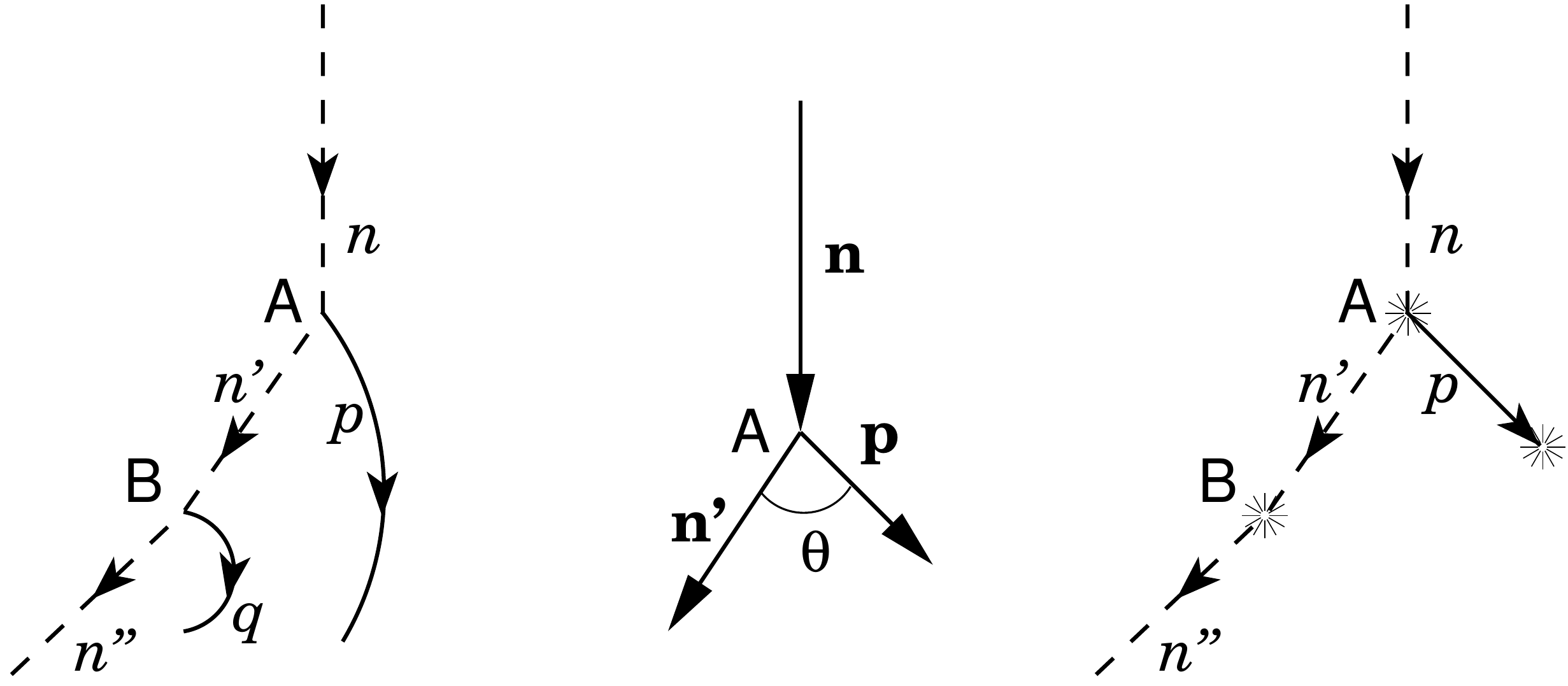}
   \caption{{Incoming neutron with 4-momentum $n$ and 3-momentum
       \vect{n} undergoing two collisions in the detector active volume.
       Left: particles trajectories in a magnetic spectrometer.  Center:
       3-momenta configuration at the first elastic collision.  Right:
       calorimeter view of the same pair of collisions.}}
   \label{fig-n-2coll}
 \end{figure}

 Figure~\ref{fig-n-2coll} shows two possibilities for neutron
 detection, in the case of a magnetic spectrometer (left) and with a
 sampling calorimeter (right).  In both cases, it is assumed that the
 3D positions of the two scattering sites A and B are measured
 together with the 4-momentum of the proton kicked off by the first
 collision.  In the case of a magnetic spectrometer, the 3-momentum
 \vect{p} of the proton is measured, whereas its total energy
 $E_{\text{p}}$ and direction are measured by the calorimeter.  The
 proton is identified by looking at the details of its energy loss and
 velocity, and the 4-momentum is reconstructed by assuming that the
 visible particle has the proton mass \Mp\ making use of the relation
 $p^2 = \Mp^2 = E_{\text{p}}^2 - \vect{p}^2$ (using units $c=1$ in
 this section).

 Conservation of 4-momentum reads $n = n' + p$ and the direction of
 the neutron after the first interaction is known from the two
 position measurements.  In order to find the initial 4-momentum $n$
 we need first to find the magnitude of $n'$.  This is obtained by taking the
 square of the 4-momentum conservation law, which gives relativistic
 invariants on both sides of the equation: $n^2 = (n' + p)^2$ gives
 \begin{equation}
   \Mn^2 = \Mn^2 + \Mp^2
         + 2(E_{\text{p}} \, E_{\text{n}'}
          - |\vect{p}| \, |\vect{n}'| \cos\theta)
 \end{equation}
 from which one finds the magnitude of the scattered neutron 3-momentum:
 \begin{equation}
   \label{eq-vec-n}
   |\vect{n}'| = \frac{\Mp^2}{2 p_L}
               + E_{\text{n}'} \, \frac{E_{\text{p}}}{p_L}
 \end{equation}
 where the longitudinal momentum $p_L = |\vect{p}| \cos\theta$ is the
 measured component of the proton 3-momentum \vect{p} along the
 direction AB of the scattered neutron. 

 The only unknown quantity in the right hand side of \eqref{eq-vec-n}
 is the energy
 \(
    E_{\text{n}'} % = \sqrt{\Mn^2+\vect{n}'^{2}}
 \)
 of the scattered neutron.  By substituting 
 \(
    E_{\text{n}'} = \gamma\Mn % c^2
 \)
 and
 \(
    |\vect{n}'| = \sqrt{\gamma^2 - 1} \, \Mn  % c^2
 \)
 in equation \eqref{eq-vec-n}, one obtains a quadratic equation for
 the Lorentz factor $\gamma$ of the scattered neutron
 \begin{equation}
   \label{eq-quadratic}
   \left( \frac{E_{\text{p}}^2}{p_L^2} - 1 \right) \gamma^2 +
   \frac{\Mp^2 \, E_{\text{p}}}{\Mn \, p_L^2} \, \gamma +
   \frac{\Mp^4}{4 \Mn^2 \, p_L^2} + 1 = 0
 \end{equation}
 with two formal roots,
 % \begin{equation}
 %   \label{eq-roots}
 %   \gamma = \frac{-\Mp^2 E_{\text{p}} \pm \sqrt{\Mp^4 p_L^2 + 4 p_L^2
 %       \Mn^2 (p_L^2 - E_{\text{p}}^2)}}{2 \Mn ( E_{\text{p}}^2 - p_L^2 )}
 % \end{equation}
 although the presence of measurement uncertainties forces one to
 employ numerical methods to find its optimal solution.
 Together with the direction AB = $\hat{\vect{n}}'$, the solution
 $\gamma$ of equation \eqref{eq-quadratic} gives the 3-momentum of the
 scattered neutron,
\(
  \vect{n}' = \sqrt{\gamma^2 - 1} \, \Mn \hat{\vect{n}}'
\),
 from which one finds the 3-momentum \vect{n} of the incoming neutron:
 \begin{equation}
   \label{eq-n-3mom}
    \vect{n} = \vect{n}' + \vect{p}
             = \sqrt{\gamma^2 - 1} \, \Mn \hat{\vect{n}}' + \vect{p} \;.
 \end{equation}
 Thus \eqref{eq-n-3mom} gives the direction of the source and allows
 us to obtain the energy of the input neutron from the relation
 $E_{\text{n}}^2 = \Mn^2 + \vect{n}^2$.

 If the total energy of the scattered neutron is also measured in the
 second collision, one gets a better estimate of the incoming neutron
 energy.  The best way to measure the total energy is to adopt a
 hadronic calorimeter as a second detection stage, because this will
 also work for events in which there is no elastic collision in the
 first place.

%%%%%%%%%%%%
\subsection{Detector design}\label{sec-detector}
%%%%%%%%%%%%

 Although no study on neutron detection has been published by the AMS
 collaboration, it is hopefully possible to select n-p elastic
 scatterings in the transition radiation detection (TRD) system of
 AMS-02 by selecting events in which a single proton track starts
 inside the TRD (i.e.\ by using the first TRD layers in
 anticoincidence) and crosses the triggering scintillators of the time
 of flight system.
 However, this only works if the efficiency of the first TRD layers is
 close to 100\% (which sounds unlikely), because proton events are
 about $10^4$ times more frequent than neutrons.  In addition, it is
 difficult to detect also the second neutron interaction inside
 AMS-02, although this is perhaps possible by tuning the selection
 criteria (this event topology would be discarded as ``noise'' in the
 standard CR selection).
 The electromagnetic calorimeter of AMS-02 can detect gamma-rays, and
 might have a small but non negligible efficiency also for neutrons.
 However, no study has been published about them yet.
 Hence at best AMS-02 has a very small acceptance for neutron events.

 The acceptance of PAMELA is about 50 times smaller than AMS-02,
 however it has a neutron detector underneath the last scintillator
 layer, to improve the identification of hadronic interactions
 \cite{Adriani_2014}.  The neutron detector is located below the S4
 scintillator and consists of 36 proportional counters, filled with
 $^3$He and surrounded by a polyethylene moderator enveloped in a thin
 cadmium layer to prevent thermal neutrons entering the detector from
 the sides and from below.
 % The counters are stacked in two planes of 18 counters, oriented along
 % the y-axis of the instrument.
 The neutron detector is active for 200 microseconds after each
 trigger, which is the time needed to thermalize neutrons produced by
 interactions of 20--180 GeV protons in the calorimeter
 \cite{Stozhkov_2005}.  As such, it seems that PAMELA is not able to
 perform a direct measurement of cosmic neutrons using the technique
 explained here.

 Scintillators work well at the MeV scale or below, at which full
 absorption is possible.  Arrays of scintillators similar to the
 CsI(TI) crystals flown on \emph{Salut-7} and \emph{Kosmos-1686}
 orbital complex \cite{bogolubov2011} might work also up to several
 GeV but this would require large volume and fine segmentation, which
 is not easy to obtain with this material.

 Liquid scintillators are also good for discriminating neutron events
 from gamma-rays.  For example, the omnidirectional neutron
 measurement by COMPTEL was performed using one D1 module, a cylinder
 with 13.8 cm radius and depth of 8 cm filled by liquid scintillator
 NE213A \cite{Morris_1995}. In addition to NE213 \cite{Lee_1998},
 other liquid scintillators are good, like the BC501A scintillator
 studied by the ICARUS collaboration \cite{Arneodo_1998}, EJ301
 (similar to BC501A), and the EJ309 scintillator, more suitable for
 environmentally difficult situations \cite{Iwanowska_2012}.  However,
 directionality is only achievable by developing a large time
 projection chamber filled by liquid scintillator, which is perhaps
 too challenging (though very interesting) for a space experiment.

 A neutron detector design which exploits double n-p elastic
 collisions is SONTRAC \cite{Wunderer1997}, consisting of a set of
 orthogonal scintillating fibres (made by organic plastic
 scintillator) on top of a BGO calorimeter.  However, its design is
 optimized to measure 20-250 MeV neutrons from solar flares.  Thus, a
 much larger detector would be needed to reach the interesting energy
 regions for indirect WIMP detection.
 \citet{Imaida_1999} also developed a neutron detector made by
 scintillating fibres and a multi-anode photomultiplier, which was
 operated in space \cite{Koga_2011,Muraki_2016}.  Like SONTRAC, it
 covers the energy range interesting for solar flares.
 \citet{moser2005} also developed a stack of plastic scintillating
 fibres, in order to detect solar flare neutrons with n-p elastic
 scattering.  They consider single-, double- and triple-scatter events
 up to about 100 MeV, but did not compute the full initial neutron
 momentum for any incoming direction, as we have done above.
 In the energy range useful for solar flares, arrays of plastic
 scintillator bars also work as fast neutron detectors
 \cite{Karsch_2001}.  At the price of a reduced granularity, they can
 cover wide surfaces.  By stacking several layers, one may obtain a
 large-acceptance neutron detector with limited directionality.

 Another interesting concept is HERO \cite{atkin2009}, although it
 is designed to detect thermalized neutrons about 100~\micro{s} after
 the trigger signal, similar to PAMELA, with the purpose of separating
 hadronic from electromagnetic showers.  Still, its capability of
 reconstructing the cascade parameters gives it directional
 capability.  What is missing is a trigger for neutral hadrons and an
 initial section of light active materials optimized for elastic
 scattering.
 \citet{Alexandrov_2001} proposed a ionization-neutron calorimeter,
 the INCA experiment, also designed to thermalize and detect neutrons.
 The primary purpose of INCA is the investigation of high-energy CR
 electrons and primary nuclei up to the knee region, hence it is a
 huge space project, with geometry close to $4\pi$, $30 \um{m}^2
 \um{sr}$ acceptance, and 10--12~t mass.  However it is not optimized
 for the measurement of the cosmic neutron flux.

 None of the detectors built or proposed so far is optimized for the
 measurement of the cosmic neutron flux over a broad energy range.  As
 we have seen, at low energies elastic scattering is the main process,
 whereas at high energies inelastic collisions dominate.  Hence one
 could design the instrument such that the face directed toward the
 source is optimized for elastic scattering, with a light active
 material, followed by a hadronic calorimeter made of a heavy
 material, for the reconstruction of hadronic showers.

 The detector should have modular design, with multiple ``towers''
 that can be tiled together.  Alternatively, independent units may be
 installed, possibly on different satellites (in this case each unit
 will be sensitive to less energetic neutrons, compared to one big
 detector).  A modular design simplifies the development, enhances the
 redundancy and fault tolerance, leaves some margin in case the
 overall mass or financial budget is downsized (in which case one
 might decide to build less units), and allows for a gradual
 commissioning phase, in case independent units may be deployed in
 sequence.

 The detector needs to be surrounded by an anticoincidence shield, in
 order to veto charged particles entering the active area from all
 directions.  The anticoincidence system can be built with plastic
 scintillators, that provide a fast response with a high light yield
 and can be produced in a variety of shapes.  Each tower must provide
 different response for neutrons and gamma-rays, in order to
 distinguish between them by means of pulse shape discrimination
 techniques.

 Symmetric towers might also be developed, with sections optimized for
 elastic scattering on opposite sides.  In this case, the detector can
 be pointed toward the Sun and simultaneously look at the opposite
 direction.  Differential measurements have the potential for showing
 a modulation, for example when Jupiter or Mars are on the same side
 as the Sun or on the opposite side.  Hence are more robust against
 systematics.  For example, Jovian and Martian neutrons produced by CR
 interactions within their atmospheres, or neutrons coming from the
 Moon and produced by CR collision with the lunar soil, may be
 important sources of background.  In principle, the reconstruction of
 the incoming neutron direction is sufficient to suppress this
 background.  However, there might be some ambiguity or the
 experimental resolution may be not sufficient in all cases, and a
 differential measurements helps in keeping several sources of
 systematics under control.

 The external segments of each tower shall provide the 3D position of
 the elastic scattering and allow for the reconstruction of the track
 left by the ejected proton.
 Organic plastic scintillators are available in the form of fibres
 with round or square cross-section, and are best suited for the first
 layers.  Alternating layers with orthogonal fibres allow for the
 reconstruction of the location of the first collision (point A in
 figure~\ref{fig-n-2coll}) within about one fibre diameter (0.25 to
 0.5 mm), and organic scintillators have a high fraction of hydrogen
 atoms, which are the best targets.  The presence of carbon nuclei as
 additional targets for the n-C elastic scattering shall be taken into
 account, but the negligible amount of heavier elements makes these
 scintillators simpler to use than inorganic scintillators.

 Liquid scintillators also provide a very good medium for elastic
 scattering, but reconstructing the proton track requires a time
 projection chamber, which is more complex (though a very attractive
 option) than a bundle of scintillating fibres.  In addition, the
 layered structure of the latter makes it easy to implement the
 anticoincidence system, by configuring the trigger system in order to
 veto events with signals detected by the outermost fibres.

 The organic scintillator fibres shall be followed by a segmented
 calorimeter, which is able to provide information on the shower shape
 in case of inelastic collision, and the position of the second
 neutron scattering (point B in figure~\ref{fig-n-2coll}) which has
 undergone an elastic collision in the external segment.  A heavy
 material with relatively short interaction length is the best choice,
 for example bismuth germanium oxide Bi$_4\,$Ge$_3\,$O$_{12}$ (or
 simply BGO).  The total thickness along the line of sight of the
 instrument should be enough to guarantee the development of a
 hadronic shower, whose containment as a function of the neutron
 energy will determine the energy and direction resolution of the
 instrument.

 Most hadronic detectors are sampling calorimeters, in which layers of
 heavy passive material (like tungsten or lead) are added to increase
 the interaction probability.  The charged secondaries ionizing the
 active material are then detected, leaving measurable signals.  In
 this case, the active material can be made of plastic scintillating
 fibres, which run most often along different directions, to provide a
 good position resolution.  This alternative design is also worth
 consideration for two reasons.  First, it has the advantage of using
 the same type of scintillator, hence also photon detector and
 associated electronics, for both light and heavy sections of each
 unit.  Second, the layered structure provides information about the
 longitudinal direction, while heavy crystals would be coupled with
 photon detectors on one side only, making it almost impossible to
 obtain a 3-dimensional reconstruction.

\begin{figure}
   \centering
   \includegraphics[width=\columnwidth]{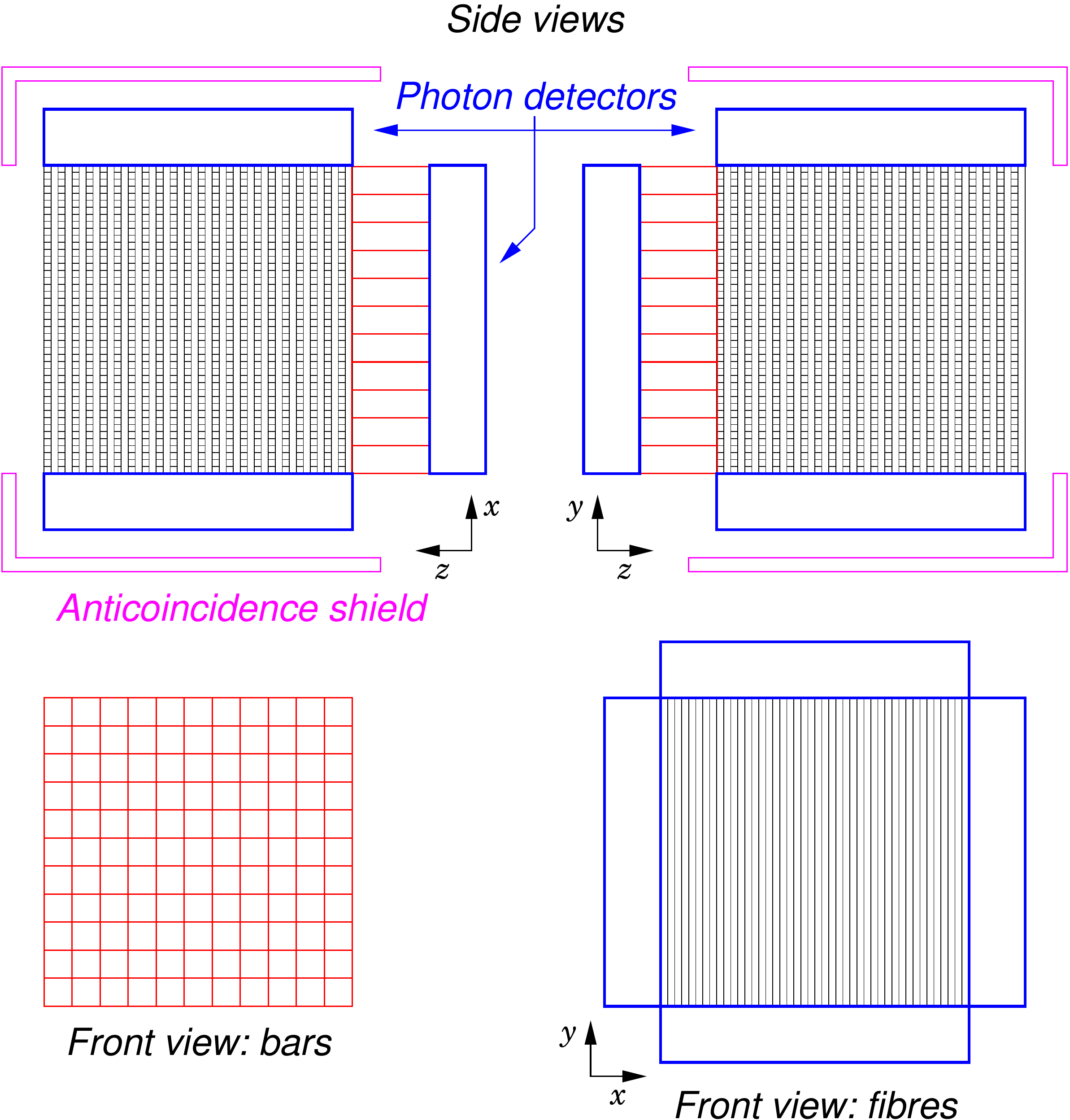}
   \caption{Detector design principle.  Scintillation fibre
     calorimeter (black) in front of crystal scintillator bars (red).
     Anticoincidence shield (magenta) surrounding the active detectors.
     Photon detectors (blue) are not shown for the anticoincidence
     scintillators.}
   \label{fig-detector}
 \end{figure}

 After having discussed the guidelines for the design of a suitable
 detector, we conclude the section with the description of a compact
 demonstrator.  We consider one unit consisting of a finely segmented
 calorimeter made of organic scintillating fibres, followed by one
 array of BGO crystals.  The unit shall be equipped with an
 anticoincidence shield as shown in figure~\ref{fig-detector}.  The
 minimal size for covering at least the energy range of solar
 flare neutrons, while keeping volume and mass as low as possible, 
 is close to a cubic $(20\um{cm})^3$ fibre scintillator (8.5 kg)
 coupled with a box $(20 \times 20 \times 5) \um{cm}^3$ containing BGO
 crystals (14.3~kg).  By including anticoincidence scintillators,
 supporting mechanics, photon detector and related electronics, this
 payload would have mass in the range 30--40 kg, well within the
 capability of a small space mission.
 Pairing back-to-back two units and reaching a mass close to 70~kg,
 one would then obtain a bidirectional instrument, which is able to
 perform differential measurements, still within the range of a small
 space mission.

 The nuclear collision length of BGO is 13.49~cm and the nuclear
 interaction length is 22.31~cm.  One problem with this material is
 the short attenuation length (1.1~cm), which limits the maximum size
 of each crystal to few centimeters.  For this reason, we consider
 here 5~cm thick crystals, which may have 2~cm width along two
 directions.  This ensures a moderate granularity, 31\% collision
 probability and 20\% interaction probability, while achieving energy
 resolution of 16\% at 511~keV and about 4000 photons from a
 minimum ionizing particle (depositing 8.92 MeV/cm) \cite{pdb2016}.
 The radiation length is only 1.118~cm in BGO, hence the detection
 efficiency for gamma-rays is very high.  This means that the same
 detector will also provide cosmic gamma-ray measurements of good
 quality, a valuable scientific bonus.  In comparison, a segmented
 hadronic calorimeter would have worse performance in gamma-ray mode.
 This is the main reason why here the inorganic crystal calorimeter is
 considered for the demonstrator mission.

 Organic scintillating fibres are available in several shapes and
 material.  To make one example, 0.25 mm diameter fibres of BCF-10 by
 St.~Gobain (essentially made of polystyrene) would provide an
 excellent spatial resolution and a high detection efficiency for
 charged particles.  The external fibres may be configured in
 anticoincidence, vetoing all events in which a charged particle
 enters from the front face and complementing the work of the
 anticoincidence shield, which flags events with tracks entering the
 detector from the side.
 Neutrons have a collision probability of 31\% in 20~cm polystyrene
 (the same as 5~cm BGO) and an interaction probability of 23\%.
 Considering only neutrons that undergo one elastic scattering in the
 fibres and are detected by BGO thanks to another recoil or to a
 hadronic interaction (51\% probability in total), this detector
 should have an efficiency of order 10\%, to be estimated more
 accurately with Monte Carlo simulations. 

 The concept described above is very similar to SONTRAC, as it offers
 several layers of scintillating fibres alternating along orthogonal
 directions followed by BGO crystals.  Hence the measured performance
 of the SONTRAC prototype \cite{Ryan1997} should provide a quite good
 reference for the demonstrator detector illustrated above, despite
 from the minor differences in geometry.
 The biggest differences from SONTRAC are the bidirectional design,
 with scintillating fibres on two opposite faces to perform
 differential measurements, and the thicker BGO section, which
 increases the sampled energy range.  In addition, this demonstrator
 may be consider one module of a bigger detector, whose main goal is
 not the study of solar flares but the search for DM induced neutrons
 with a high sensitivity.
 The actual instrument configuration (like dimensions, precise layout,
 photodetectors, etc.)\ shall be studied by means of Monte Carlo
 simulations, together with alternative designs (for example a
 sampling calorimeter with W layers alternating with scintillating
 fibres), in order to optimize the performance.

%%%%%%%%%%%%%%%%%%%%%%%%%%%%%%%%%%%%%%%%%%%%%%%%%%%%%%%%%
\section{Summary}\label{sec-summary}
%%%%%%%%%%%%%%%%%%%%%%%%%%%%%%%%%%%%%%%%%%%%%%%%%%%%%%%%%

 Cosmic neutrons can only reach the Earth if their energy is high
 enough that the relativistic time expansion effect ensures that they
 do not decay during their travel with high probability.
 For example, neutrons from the Sun need to have 20 MeV or higher
 energy.  Neutrons from the closest supernova remnants can reach us
 only at PeV energies or above, in the knee region of the CR spectrum.
 Neutrons from the Galaxy center should have ultra high energy,
 affecting the CR ankle region.

 Between solar neutrons, which can be emitted during a solar flare and
 are well correlated in time with the X-ray emission, and the other
 expected sources of cosmic neutrons, located well outside the solar
 system, there is a wide energy range in which the diffuse cosmic
 neutron flux should be not very dissimilar from the cosmic antiproton
 flux, which amounts to about $10^{-4}$ of CR protons.
 This is smaller than the secondary neutrons produced by CR
 interactions in the Earth atmosphere, at least over the energy range
 at which they have been already measured.  Hence a directional
 detector operating in orbit is required, in order to get rid of this
 background.

 A detector pointing to the Sun and able to measure the neutron flux
 also from the opposite side is proposed.  It is also suggested to
 adopt several layers of orthogonal fibres of organic scintillators at
 the entrance sides, in order to study the kinematics of elastic
 collisions below few GeV, enclosing a heavier core in which inelastic
 interactions can produce hadronic showers.  The central section may
 consist of BGO crystals, or be designed as a sampling calorimeter.
 This way, the neutron energy spectrum can be obtained at low energies
 by exploiting the kinematics of elastic n-p scattering as explained
 in section~\ref{sec-elastic}, and at high energies by adopting a
 hadronic calorimeter.

 Charge particles shall be vetoed by an anticoincidence system, made
 with plastic scintillators surrounding the neutron detector.
 Gamma rays will be distinguished by means of pulse shape
 discrimination and by the shower development in the calorimeter.
 Finally, a modular design is possible, in which identical towers are
 tiled together.

 Sun pointing is preferred for two reasons.  In addition to the
 possibility of detecting solar neutrons, which is important in
 understanding the acceleration mechanisms during solar flares, WIMP
 annihilations should be enhanced by the gravitational field of the
 Sun.  This should increase the flux of neutrons entering from the Sun
 side, with respect to those entering from the dark side.  In
 addition, the symmetric design is also motivated by the fact that
 differential measurements are robust against several sources of
 systematic effects.  For example, the flux of secondary Jovian and
 Martian neutrons will show time variations according to their
 position.
 Any source of antiproton-proton signals will also produce
 antineutron-neutron pairs.  Hence looking for an excess of cosmic
 neutrons is a very effective and low-background search for dark
 matter particles.

\section*{Acknowledgements} 

I'm grateful to Karel Kudela for the comments on the first draft of
this paper and for pointing me to his excellent review of solar flare
neutrons \cite{Kudela09}.  He also kindly provided me with valuable
references, which I had no access to.
Next, I wish to thank Marco Cirelli, who has verified the correctness
of the assumption of similar production rates for neutrons and
antiprotons and introduced me to PPPC 4 DM ID \cite{Cirelli2011}.
I also want to express my gratitude to Doojin Kim, for the
discussion on the implications of DM ``transporting'' mechanisms
\cite{Kim2017} on cosmic neutrons.
Finally I really appreciated the suggestions of the anonymous
reviewer, which helped me to significantly improve this paper.

%%%%%%%%%%%%%%%%%%%%%%%%%%%%%%%%%%%%%%%%%%%%%%%%%%%

\bibliography{neutronbib}

\end{document}